\documentclass [12pt] {report}
\usepackage{amssymb}
\newcommand {\eqdef} {\stackrel{\rm def}{=}}
\newcommand {\D}[2] {\displaystyle\frac{\partial{#1}}{\partial{#2}}}

\newcommand {\Dd}[3] {\displaystyle\frac{\partial^2{#1}}{\partial{#2}\partial{#3}}}
\newcommand {\al} {\alpha}

\newcommand {\ga} {\gamma}

\newcommand {\si} {\sigma}

\newcommand {\de} {\delta}
\newcommand {\prtl} {\partial}
\newcommand {\fr} {\displaystyle\frac}

\newcommand {\be} {\begin{equation}}
\newcommand {\ee} {\end{equation}}
\newcommand {\ba} {\begin{array}}
\newcommand {\ea} {\end{array}}
\newcommand {\bp} {\begin{picture}}
\newcommand {\ep} {\end{picture}}
\newcommand {\bc} {\begin{center}}
\newcommand {\ec} {\end{center}}
\newcommand {\bt} {\begin{tabular}}
\newcommand {\et} {\end{tabular}}
\newcommand {\lf} {\left}
\newcommand {\rg} {\right}

\newcommand {\cF} {{\cal F}}

\newcommand {\cR} {{\cal R}}
\newcommand {\cS} {{\cal S}}

\newcommand {\iy} {\infty}
\newcommand {\ses} {\medskip}

\newcommand {\e} {\mathop{\rm e}\nolimits}

\newcommand {\arccot} {\mathop{\rm arccot}\nolimits}

\newcommand {\bR} {{\bf R}}

\newcommand {\g}  {\stackrel{g\to -g}{\Longleftrightarrow}}

\newcommand {\cE} {{\cal E}}
\newcommand {\cP} {{\cal P}}
\newcommand {\bibit} {\bibitem}
\newcommand {\nin} {\noindent}

\renewcommand{\theequation}{\arabic{sctn}.\arabic{equation}}
\def\2#1#2#3{{#1}_{#2}\hspace{0pt}^{#3}}
\def\3#1#2#3#4{{#1}_{#2}\hspace{0pt}^{#3}\hspace{0pt}_{#4}}
\newcounter{sctn}
\def\sec#1.#2\par{\setcounter{sctn}{#1}\setcounter{equation}{0}
                  \noindent{\bf\boldmath#1.#2}\bigskip\par}

\begin {document}

\begin {titlepage}

\vspace{0.1in}

\begin{center}
{\Large
Finsleroid-Space Supplemented by Angle
}\\
\end{center}

\vspace{0.3in}

\begin{center}

\vspace{.15in}
{\large G.S. Asanov\\}
\vspace{.25in}
{\it Division of Theoretical Physics, Moscow State University\\
119992 Moscow, Russia\\
(e-mail: asanov@newmail.ru)}
\vspace{.05in}

\end{center}

\begin{abstract}

Our previous exploration
of the
$\cE_g^{PD}$-geometry
has shown that the field is promising. Namely,
the
$\cE_g^{PD}$-approach
is amenable to development of novel
trends in relativistic and metric differential geometry
and can particularly be effective in context of the Finslerian
or Minkowskian Geometries.
The main point of the present paper is the tenet that the
$\cE_g^{PD}$-space-associated
one-vector
Finslerian metric function
admits in quite a natural way
an attractive  two-vector extension,
thereby giving rise to angle and scalar product.
The underlying idea is to derive the angular measure from
the solutions to the geodesic equation, which prove to be obtainable
in an  explicit simple form.
The respective investigation is presented in Part I.
Part II serves as an extended Addendum enclosing the material which is
primary for the
$\cE_g^{PD}$-space.
The Finsleroid, instead of the unit sphere,
is taken now as carrier proper of the spherical image.
The indicatrix is, of course, our primary tool.

\end{abstract}

\end{titlepage}

\vskip 1cm

{\bf  INTRODUCTION}
\bigskip

Various known   attempts  to introduce the concept of angle in
the Minkowskian or Finslerian spaces
[1-8]
were steadily encountered with  drawback positions:

\ses

"Therefore no particular angular measure can be entirely natural in
Minkowski geometry. This is evidenced by the innumerable attempts to define
such a measure, none of which found general acceptance``.
(Busemann [2], p. 279.)

\ses

"Unfortunately, there exists a number of distinct invariants
in a Minkowskian space all of which
reduce to the same classical
euclidean invariant if
the Minkowskian space
degenerates into
a euclidean space.
Consequently,
distinct definitions of
the trigonometric functions and of  angles
have appeared in the literature concerning
 Minkowskian and
Finsler spaces``. (Rund [3], p. 26)

\ses

The fact that the attempts have never been unambiguous seems to be due
to a lack of the proper tools.
The opinion was taken for granted that
the angle  ought to be defined or constructed
in terms
of the basic
Finslerian metric tensor
(and whence ought to be explicated from the initial Finslerian metric function).
Let us doubt  the opinion from the very beginning.
Instead, we would like to raise alternatively the tenet that
the angle
 is a concomitant of the geodesics
(and  not  of the metric proper).
The angle is determined by two vectors (instead of one vector in case of the
length) and actually
implies using a due extension of the Finslerian metric function
to a two-vector metric function
(the scalar product).
Below, we apply this tenet to study the
$\cE_g^{PD}$-spaces
[9-12]
in which certainly
one needs to use not only length but also angle and  scalar product.

\ses

Accordingly, in  Part I,
we first deal with the geodesic equation
(Sec. 1.1).
Remarkably, the equation admits a simple and
explicit general solution.
 After that, the angle between two vectors is explicated.
It is commonly expected that the angular measure should be additive for
angles with the same vertex.
Remarkably,
the angle found is a factor of the euclidean angle and, therefore, is additive.
The Cosine Theorem remains valid
if the euclidean angle is replaced by the angle found.
The respective scalar product is obtained.

In Sec. 1.2, we introduce the associated two-vector metric tensor
and demonstrate that at the equality of vectors the tensor reduces
exactly to the one-vector Finslerian metric tensor of the
$\cE_g^{PD}$-space. The concomitant two-vector metric tensor is
given by the components (2.2). The orthonormal frame thereto is
also found in a lucid explicit form. After that, in Sec. 1.3, the
possibility of converting the theory into the co-approach is
presented, and in Sec. 1.4 the $\cE_g^{PD}$-extension of the
parallelogram law of vector addition is derived; it occurs
possible to find the sum vector and the difference vector in a
nearest approximation. Part I ends with Sec. 1.5 in which we
return the treatment from the auxiliary quasi-euclidean framework
to the primary $\cE_g^{PD}$-approach.

Part II reviews the fundamental ingredients of the $\cE_g^{PD}$-space
in great detail.

\ses
\ses

{\bf\large PART I:}

\ses
\ses

{\bf\large
$\cE_g^{PD}$-SPACE
GEODESICS, ANGLE, SCALAR PRODUCT}
\bigskip

\ses
\ses

\setcounter{sctn}{1}
\setcounter{equation}{0}
{\nin\bf 1.1.
Derivation of geodesics
and angle
in
associated
 quasi-euclidean space
}

\ses
\ses

For the space under study, the geodesics should be obtained as solutions to the equation
\be
\fr
{d^2R^p}
{ds^2} +\3Cqpr(g;R)
\D{R^q}{ds}
\D{R^r}{ds}
=0
\ee
which coefficients
$\3Cpqr$
are given by the list placed the end of Sec. 2.2 of Part II.
To avoid complications of calculations involved, it proves convenient to
transfer the consideration in the quasi-euclidean approach
(see Sec. 2.3 of Part II). Accordingly, we put
\be
\sqrt{
g_{pq}(g;R)dR^pdR^q}
=
\sqrt{
n_{pq}(g;t)dt^pdt^q}
\ee
and
\be
R^p(s)=
\mu^p(g;t^r(s))
\ee
together with
\be
\fr
{dR^p(s)}
{ds}
=
\mu^p_q(g;t^r(s))
\fr
{dt^q(s)}
{ds},
\ee
where
$
\mu^p(g;t^r)
$
and
$
\mu^p_q(g;t^r)
$
are the coefficients given, respectively by Eqs. (3.14) and
(3.38)-(3.40) of Part II.
Let a curve $C$:\,
$t^p=t^p(s)$
be given
in
the
quasi-euclidean space,
with
{\it the
arc-length parameter} $s$ along the curve
being defined by the help of the differential
\be
ds=\sqrt{
n_{pq}(g;t)dt^pdt^q},
\ee
where
$
n_{pq}(g;t)
$
is the associated quasi-euclidean metric tensor
given by
Eq. (3.49) in Part II.
Respectively,
{\it
the tangent vectors
}
\be
u^p=\fr{dt^p}{ds}
\ee
to the curve
are unit,
 in the sense that
\be
n_{pq}(g;t)u^pu^q=1.
\ee
Since
$L_p=\partial S/\partial t^p$,
we have
\be
L_pu^p=\fr{dS}{ds}.
\ee
Here,
$
S^2(t)=
n_{pq}(g;t)t^pt^q
=
r_{pq}t^pt^q
$
(see Eq. (3.46) in Part II).
Using
Eq. (4.16) of Part II
leads through well-known arguments (see, e.g., [13-14])
to the following
equation of geodesics in the quasi-euclidean space:
\be
\fr{d^2{\bf t}}{ds^2}=
\fr14G^2\fr{{\bf t}}{S^2}H_{pq}u^pu^q,
\ee
\ses
where
$H_{pq}=h^2(n_{pq}-L_pL_q)
$
(see Eq. (4.4) in Part II)
and
$
{\bf t}=\{t^p\}.
$
We obtain
\be
\fr{d^2{\bf t}}{ds^2}=
\fr14g^2\fr{{\bf t}}{S^2}
\lf(
1-(\fr
{dS}{ds})^2
\rg)
=
\fr14g^2(a^2-b^2)\fr{{\bf t}}{S^4}
\ee
\ses
and
\be
\fr{d^2{\bf t}}{ds^2}=
\fr14g^2(a^2-b^2)\fr{{\bf t}}{S^4}
\ee
with
\be
S^2(s)=a^2+2bs+s^2,
\ee
where $a$ and $b$ are two constants of integration.

If we put
\be
S
(\Delta s)=
\sqrt{
a^2+2b
\Delta s
+
(\Delta s)^2
}
\ee
and
\be
{\bf t}_1
=
{\bf t}(0),
\qquad
{\bf t}_2
=
{\bf t}
(\Delta s),
\ee
then
we get
\be
a=
\sqrt{
(
{\bf t}_1
{\bf t}_1
)
}
\ee
and
\be
S
(\Delta s)=
\sqrt{
({\bf t}_2
{\bf t}_2
)
}
\ee
\ses
together with
\be
(
{\bf t}_1
{\bf t}_2
)
=
 a
S
(\Delta s)
\cos
\Bigl[
h\arctan\fr{
\sqrt{a^2-b^2}
\,
\Delta s
}{a^2+b
\Delta s
}
\Bigr].
\ee
Here,
$
{\bf t}_1
$
and
$
{\bf t}_2
$
are two vectors with the fixed origin
$O$; they point to the beginning of the
geodesic and to the end of the geodesic, respectively.
The notation
parenthesis couple
$
(..)
$
is used
for the euclidean scalar product, so that
$
(
{\bf t}_1
{\bf t}_1
)
= r_{pq}t_1^pt_1^q$,
$
(
{\bf t}_1
{\bf t}_2
)
= r_{pq}t_1^pt_2^q$, and
$r_{pq}$
 is a euclidean metric tensor;
$r_{pq}=\de_{pq}$
in case of orthogonal basis; $\de$ stands for the Kronecker symbol.
From (1.15)-(1.17) it directly follows that
\be
\fr{
\sqrt{a^2-b^2}
\,
\Delta s
}
{a^2+b
\Delta s
}
=
\tan
\Bigl[
\fr1h\arccos
\fr{
(
{\bf t}_1
{\bf t}_2
)}
{
\sqrt{(
{\bf t}_1
{\bf t}_1
)}
\,
\sqrt{(
{\bf t}_2
{\bf t}_2
)}
}
\Bigr].
\ee
The  equality (1.18)  suggests the idea to introduce

\ses

{\bf DEFINITION}. {\it
The
$\cE_g^{PD}$-associated angle
}
is given by
\be
\al\eqdef
\fr1h\arccos
\fr{
(
{\bf t}_1
{\bf t}_2
)}
{
\sqrt{(
{\bf t}_1
{\bf t}_1
)}
\,
\sqrt{(
{\bf t}_2
{\bf t}_2
)}
},
\ee
 so that
\be
\al
=\fr1h\al_{euclidean}.
\ee

\ses

Such an angle is obviously
{\it
additive}:
\be
\al(
{\bf t}_1,
{\bf t}_3
)
=
\al(
{\bf t}_1,
{\bf t}_2
)
+
\al(
{\bf t}_2,
{\bf t}_3
).
\ee
Also,
\be
\al(
{\bf t},
{\bf t}
)=0.
\ee

With the angle (1.19), we ought to propose

\ses

{\bf DEFINITION}.
Given two vectors
$
{\bf t}_1
$
and
$
{\bf t}_2,
$
we say that the vectors are
{\it
$\cE_g^{PD}$-perpendicular,
}
if
\be
\cos
\lf(
\al
(
{\bf t}_1,
{\bf t}_2)
\rg)
=0.
\ee

Since the vanishing (1.23) implies \be \al_{ quasi-euclidean} (
{\bf t}_1, {\bf t}_2) = \fr{\pi}2, \ee in view of 1.20) we ought
to conclude that \be \al_{euclidean} ( {\bf t}_1, {\bf t}_2) =
\fr{\pi}2h\le \fr{\pi}2. \ee Therefore, vectors perpendicular in
the quasi-euclidean sense proper look like acute vectors as
observed from associated euclidean standpoint.

With the equality
\be
(
\sqrt{a^2-b^2}
\,
\Delta s)^2
+(a^2+b\Delta s)^2\equiv
a^2S^2(\Delta s),
\ee
we
also establish
the relations
\be
\sqrt{a^2-b^2}
\,
\Delta s=aS(\Delta s)
\sin
\al
\ee
and
\ses\\
\be
a^2+b\Delta s=
aS(\Delta s)
\cos
\al.
\ee
They entail the equality
\be
\fr b{\sqrt{a^2-b^2}}
=
\fr{
S(\Delta s)
\cos
\al
-a}{
S(\Delta s)
\sin
\al
}
\ee
from which the quantity $b$ can be explicated.

Thus
{\it
each member of the involved set
$\{
a, b, \Delta s, S(\Delta s)\}
$
can be explicitly expressed through the input vectors
$
{\bf t}_1
$
and
$
{\bf t}_2.
$
}
For many cases it is worth rewriting the equality (1.24) as
\be
S^2(\Delta s)=
(\Delta s)^2-a^2+
2(a^2+b
\Delta s
).
\ee

Thus we have arrived at the following substantive items:

\ses

\nin
{\it
The
$\cE_g^{PD}$-Case
Cosine Theorem
}
\ses\\
\be
(\Delta s)^2
=
S^2(\Delta s)
+a^2-
2aS(\Delta s)
\cos
\al\,;
\ee

\ses

\nin
{\it
The
$\cE_g^{PD}$-Case
Two-Point Length
}
\ses\\
\be
(\Delta s)^2
=
({\bf t}_1
{\bf t}_1
)
+
({\bf t}_2
{\bf t}_2
)
-2
\sqrt
{
({\bf t}_1
{\bf t}_1
)
}
\,
\sqrt
{
({\bf t}_2
{\bf t}_2
)
}
\cos
\al\,;
\ee

\ses

\nin
{\it
The
$\cE_g^{PD}$-Case
Scalar Product
}
\ses\\
\be
<{\bf t}_1,
{\bf t}_2
>
=
\sqrt
{
({\bf t}_1
{\bf t}_1
)
}
\,
\sqrt
{
({\bf t}_2
{\bf t}_2
)
}
\cos
\al\,;
\ee

\ses

\nin
{\it
The
$\cE_g^{PD}$-Case
Perpendicularity
}
\ses\\
\be
<{\bf t}_1,
{\bf t}_2
>
=
\sqrt
{
({\bf t}_1
{\bf t}_1)
}
\,
\sqrt
{
({\bf t}_2
{\bf t}_2)
}.
\ee

\ses

The identification
\ses\\
\be
|
{\bf t}_2
\ominus
{\bf t}_1
|^2
=(\Delta s)^2
\ee
yields another lucid representation
\be
|
{\bf t}_2
\ominus
{\bf t}_1
|^2
=
({\bf t}_1
{\bf t}_1
)
+
({\bf t}_2
{\bf t}_2
)
-2
\sqrt
{
({\bf t}_1
{\bf t}_1
)
}
\,
\sqrt
{
({\bf t}_2
{\bf t}_2
)
}
\cos
\al\,.
\ee

The consideration can be completed by

\ses

{\bf THEOREM}.
{\it
A general solution to the geodesic equation {\rm (1.11)}
can explicitly be found as follows:
\ses
\ses\\
$$
{\bf t}(s)=
$$
\ses
\ses
\be
=
\fr {S(s)}{
 a
}
\fr
{\sin
\Bigl[
h\arctan\fr{
\sqrt{a^2-b^2}
\,
(\Delta s-s)
}{a^2+b
\Delta s
+(b+
\Delta s
)
 s
}
\Bigr]
}{
\sin
\Bigl[
h\arctan\fr{
\sqrt{a^2-b^2}
\,
\Delta s
}{a^2+b
\Delta s
}
\Bigr]
}
\,
{\bf t}_1
+
\fr{ S(s)}{
S
(\Delta s)
}
\fr
{\sin
\Bigl[
h\arctan\fr{
\sqrt{a^2-b^2}
\,
s
}{a^2+bs
}
\Bigr]
}{
\sin
\Bigl[
h\arctan\fr{
\sqrt{a^2-b^2}
\,
\Delta s
}{a^2+b
\Delta s
}
\Bigr]
}
\,
{\bf t}_2.
\ee
}

\ses

The euclidean limit proper is
\ses\\
$$
{{\bf t}(s)}
_{\Bigl|_
{g=0}
\Bigr{.}}
=
\fr{
(\Delta s-s)
{\bf t}_1
+
s{\bf t}_2
}{
\Delta s
}
=
{\bf t}_1+
(
{\bf t}_2-
{\bf t}_1
)\fr s{\Delta s},
$$
so that the geodesics become straight.
From (1.35) the equality
\ses
\be
(
{\bf t}(s)
{\bf t}(s)
)
=S^2(s)
\ee
follows, in agreement with (1.12).
Since the general solution (1.35) is such that the right-hand side
is spanned by two fixed vectors,
$
{\bf t}_1
$
and
$
{\bf t}_2,
$
we are entitled concluding that
{\it
the geodesics under study are plane curves.
}

Calculating the first derivative yields
simply
the formula
\ses\\
$$
\fr{d{\bf t}
}
{ds}
(s)=
\fr{b+s}{S^2(s)}
{\bf t}
\,
-
\fr{
\sqrt{a^2-b^2}
\,
h
}
{ aS(s)}
\fr
{\cos
\Bigl[
h\arctan\fr{
\sqrt{a^2-b^2}
\,
(\Delta s-s)
}
{a^2+b
\Delta s
+(b+
\Delta s
)
 s
}
\Bigr]
}{
\sin
\Bigl[
h\arctan\fr{
\sqrt{a^2-b^2}
\,
\Delta s
}{a^2+b
\Delta s
}
\Bigr]
}
\,
{\bf t}_1
$$
\ses
\ses
\ses
\ses
\ses
\ses
\be
+
\fr{
\sqrt{a^2-b^2}
\,
h}{
S(s)
S
(\Delta s)
}
\fr
{\cos
\Bigl[
h\arctan\fr{
\sqrt{a^2-b^2}
\,
s
}{a^2+bs
}
\Bigr]
}{
\sin
\Bigl[
h\arctan\fr{
\sqrt{a^2-b^2}
\,
\Delta s
}{a^2+b
\Delta s
}
\Bigr]
}
\,
{\bf t}_2,
\ee
in which the representation (1.35) should be inserted.
The right-hand part here is such that
$$
{\bf t}(s)
\lf(
\fr{d{\bf t}
}
{ds}
(s)-
\fr{b+s}{S^2(s)}
{\bf t}
\rg)
=0,
$$
from which observation the useful equality
\be
{\bf t}(s)
\fr{d{\bf t}
}
{ds}
(s)=b+s
\ee
ensues.

For the vectors
\be
{\bf b}_1
\eqdef
\fr12
\D{
|{\bf t}_2\ominus
{\bf t}_1|^2
}
{
{\bf t}_1
},
\qquad
{\bf b}_2
\eqdef
\fr12
\D{
|{\bf t}_2\ominus
{\bf t}_1|^2
}
{
{\bf t}_2
},
\ee
we can obtain the simple representations
\ses
\ses\\
\be
{\bf b}_1
=
{\bf t}_1
-
\fr
{
{\bf t}_1
}
{
\sqrt
{
(
{\bf t}_1
{\bf t}_1
)
}
}
\sqrt
{
(
{\bf t}_2
{\bf t}_2
)
}\,
\cos
\al
-
\fr{
\sqrt{(
{\bf t}_2
{\bf t}_2
)}
}
{
h
\sqrt{(
{\bf t}_1
{\bf t}_1
)}
}
{\bf d}_1
\sin
\al
\ee
and
\be
{\bf b}_2
=
{\bf t}_2
-
\fr
{
{\bf t}_2
}
{
\sqrt
{
(
{\bf t}_2
{\bf t}_2
)
}
}
\sqrt
{
(
{\bf t}_1
{\bf t}_1
)
}\,
\cos
\al
-
\fr{
\sqrt{(
{\bf t}_1
{\bf t}_1
)}
}
{
h
\sqrt{(
{\bf t}_2
{\bf t}_2
)}
}
{\bf d}_2
\sin
\al,
\ee
\ses
where
the convenient vectors\ses\\
\be
{\bf d}_1
=
\fr{
(
{\bf t}_1
{\bf t}_1
)
{\bf t}_2
-
(
{\bf t}_1
{\bf t}_2
)
{\bf t}_1
}
{
\sqrt
{(
{\bf t}_1
{\bf t}_1
)
(
{\bf t}_2
{\bf t}_2
)
-
(
{\bf t}_1
{\bf t}_2
)^2
}
},
\qquad
{\bf d}_2
=
\fr{
(
{\bf t}_2
{\bf t}_2
)
{\bf t}_1
-
(
{\bf t}_1
{\bf t}_2
)
{\bf t}_2
}
{
\sqrt
{
(
{\bf t}_1
{\bf t}_1
)
(
{\bf t}_2
{\bf t}_2
)
-
(
{\bf t}_1
{\bf t}_2
)^2
}
}
\ee
have been introduced.
It can readily be verified that
\be
(
{\bf t}_1
{\bf d}_1
)=0,
\qquad
(
{\bf t}_2
{\bf d}_2
)=0,
\ee
\ses
\ses
\be
(
{\bf d}_1
{\bf d}_2
)=-
(
{\bf t}_1
{\bf t}_2
),
\qquad
(
{\bf d}_1
{\bf d}_1
)=
(
{\bf t}_1
{\bf t}_1
),
\qquad
(
{\bf d}_2
{\bf d}_2
)=
(
{\bf t}_2
{\bf t}_2
),
\ee
\ses
\ses
\be
(
{\bf d}_1
{\bf t}_2
)
=
(
{\bf t}_1
{\bf d}_2
)
=
\sqrt
{
(
{\bf t}_1
{\bf t}_1
)
(
{\bf t}_2
{\bf t}_2
)
-
(
{\bf t}_1
{\bf t}_2
)^2
},
\ee
\ses
and
\ses\\
\be
{\bf t}_1
{\bf b}_1
+
{\bf t}_2
{\bf b}_2
=
2
|{\bf t}_2\ominus
{\bf t}_1|^2,
\ee
\ses
together with
\be
\lim_{
{\bf t}_2\to
{\bf t}_1
}
\Bigl\{
{\bf b}_1
\Bigl\}=
\lim_{
{\bf t}_2\to
{\bf t}_1
}
\Bigl\{
{\bf b}_2
\Bigl\}=0.
\ee
\ses
\ses
For the products of the vectors (1.35) and (1.36) we obtain
\ses
$$
(
{\bf b}_1
{\bf b}_1
)
=
(
{\bf t}_1
{\bf t}_1
)
+
(
{\bf t}_2
{\bf t}_2
)
-2
\sqrt
{
(
{\bf t}_1
{\bf t}_1
)
}
\sqrt
{
(
{\bf t}_2
{\bf t}_2
)
}
\,
\cos
\al
+
\lf(\fr1{h^2}-1\rg)
(
{\bf t}_2
{\bf t}_2
)
\sin^2
\al,
$$
\ses
\ses
\ses
$$
(
{\bf b}_2
{\bf b}_2
)
=
(
{\bf t}_2
{\bf t}_2
)
+
(
{\bf t}_1
{\bf t}_1
)
-2
\sqrt
{
(
{\bf t}_2
{\bf t}_2
)
}
\sqrt
{
(
{\bf t}_1
{\bf t}_1
)
}
\,
\cos
\al
+
\lf(\fr1{h^2}-1\rg)
(
{\bf t}_1
{\bf t}_1
)
\sin^2
\al,
$$
\ses
and
\ses
$$
(
{\bf b}_1
{\bf b}_2
)
=
-
\Bigl[
\lf(
\fr
{
\sqrt{
(
{\bf t}_1
{\bf t}_1
)
}
}
{
\sqrt
{
(
{\bf t}_2
{\bf t}_2
)
}
}
+
\fr
{
\sqrt{
(
{\bf t}_2
{\bf t}_2
)
}
}
{
\sqrt
{
(
{\bf t}_1
{\bf t}_1
)
}
}
-2\cos\al
\rg)
\cos
\al
+
\lf(\fr1{h^2}-1\rg)\sin^2\al
\Bigr]
(
{\bf t}_1
{\bf t}_2
)
$$
\ses
\ses
$$
-
\fr1h
\lf(
\fr
{
\sqrt{
(
{\bf t}_1
{\bf t}_1
)
}
}
{
\sqrt
{
(
{\bf t}_2
{\bf t}_2
)
}
}
+
\fr
{
\sqrt{
(
{\bf t}_2
{\bf t}_2
)
}
}
{
\sqrt
{
(
{\bf t}_1
{\bf t}_1
)
}
}
-2\cos\al
\rg)
\sqrt{
(
{\bf t}_1
{\bf t}_1
)
(
{\bf t}_2
{\bf t}_2
)
-
(
{\bf t}_1
{\bf t}_2
)^2
}
\,
\sin
\al.
$$

The following limit
\be
\lim_{
{\bf t}_2\to
{\bf t}_1
}
\Bigl\{
\fr{
(
{\bf t}_1
{\bf t}_1
)
(
{\bf t}_2
{\bf t}_2
)
}
{
h
\sqrt{(
{\bf t}_1
{\bf t}_1
)}
\sqrt{(
{\bf t}_2
{\bf t}_2
)}
}
\fr{
\sin
\Bigl[
\fr1h\arccos
\fr{
(
{\bf t}_1
{\bf t}_2
)}
{
\sqrt{(
{\bf t}_1
{\bf t}_1
)}
\,
\sqrt{(
{\bf t}_2
{\bf t}_2
)}
}
\Bigr]
}
{
\sqrt{
(
{\bf t}_1
{\bf t}_1
)
(
{\bf t}_2
{\bf t}_2
)
-
(
{\bf t}_1
{\bf t}_2
)^2
}
}
\Bigl\}
=
\fr1{h^2}
\ee
is important to note.

\ses
\ses

\setcounter{sctn}{2}
\setcounter{equation}{0}
{\nin\bf 1.2.
The two-vector metric tensor and frame
in quasi-euclidean space
}

\ses
\ses

Now we are able to introduce
{\it
the quasi-euclidean two-vector metric tensor}
$
n(g;
{\bf t}_1,
{\bf t}_2
)
$
by the components
\be
n_{pq}(g;
{\bf t}_1,
{\bf t}_2
)
\eqdef
\Dd{
<
{\bf t}_1,
{\bf t}_2>
}
{
t^q_2
}
{
 t^p_1
}
=
-\fr12
\Dd{
|{\bf t}_2\ominus
{\bf t}_1|^2
}
{
t^q_2
}
{
 t^p_1
}.
\ee
Straightforward calculations
(on the basis of (1.32) and (1.19))
show that
\ses\\
$$
n_{pq}(g;
{\bf t}_1,
{\bf t}_2
)
=
\fr{
(
{\bf t}_1
{\bf t}_1
)
(
{\bf t}_2
{\bf t}_2
)
}
{
h
\sqrt{(
{\bf t}_1
{\bf t}_1
)}
\sqrt{(
{\bf t}_2
{\bf t}_2
)}
}
\fr{
\sin
\al
}
{
\sqrt{
(
{\bf t}_1
{\bf t}_1
)
(
{\bf t}_2
{\bf t}_2
)
-
(
{\bf t}_1
{\bf t}_2
)^2
}
}
r_{pq}
$$
\ses
\ses
\be
+
\fr{
1
}
{
\sqrt
{
(
{\bf t}_1
{\bf t}_1
)
}
\sqrt
{
(
{\bf t}_2
{\bf t}_2
)
}
}
A_1
 t_{1p}
t_{2q}
-
\fr{
1
}
{
h
\sqrt{(
{\bf t}_1
{\bf t}_1
)}
\sqrt{(
{\bf t}_2
{\bf t}_2
)}
}
A_2
d_{1p}
 d_{2q},
\ee
\ses\\
where
\be
A_1=
\cos
\al
-
\fr1h
(
{\bf t}_1
{\bf t}_2
)
\fr{
\sin
\al
}
{
\sqrt{
(
{\bf t}_1
{\bf t}_1
)
(
{\bf t}_2
{\bf t}_2
)
-
(
{\bf t}_1
{\bf t}_2
)^2
}
}
\ee
and
\be
A_2=
\fr1h
\cos
\al
-
(
{\bf t}_1
{\bf t}_2
)
\fr{
\sin
\al
}
{
\sqrt{
(
{\bf t}_1
{\bf t}_1
)
(
{\bf t}_2
{\bf t}_2
)
-
(
{\bf t}_1
{\bf t}_2
)^2
}
}.
\ee
\ses
\ses
For the determinant of the tensor (2.2) we find simply
\ses\\
\be
\det\lf(
n_{pq}(g;
{\bf t}_1,
{\bf t}_2
)
\rg)
=\lf(
\fr
{
\sqrt{({\bf t}_1{\bf t}_1)({\bf t}_2{\bf t}_2)}\sin\al
}
{
\sqrt
{
({\bf t}_1{\bf t}_1)({\bf t}_2{\bf t}_2)-({\bf t}_1{\bf t}_2)^2}
}
\rg)^{N-2}h^{-N}
\det\lf(r_{ab}\rg).
\ee

\ses

Owing to (1.43), we can establish
 the following fundamental identification:
\be
\lim_{
{\bf t}_2\to
{\bf t}_1=
{\bf t}
}
\Bigl\{
n_{pq}(g;
{\bf t}_1,
{\bf t}_2
)
\Bigl\}=
n_{pq}(g;t),
\ee
\ses
\ses
\ses
where
$
n_{pq}(g;
{\bf t}
)
$
is the quasi-euclidean metric tensor (see (3.49) in Part II).

Differentiating (2.2)  results in
$$
\D
{
n_{pq}(g;
{\bf t}_1,
{\bf t}_2
)
}
{
 t_1^s
}
=
-\fr1h
\fr
{
\sqrt
{
(
{\bf t}_2
{\bf t}_2
)
}
}
{
\sqrt
{
(
{\bf t}_1
{\bf t}_1
)
}
u
(
{\bf t}_1,
{\bf t}_2
)
}
A_2d_{1s}r_{pq}
$$
\ses
\ses
$$
+
\fr
1
{
\sqrt{
(
{\bf t}_1
{\bf t}_1
)
}
\sqrt{
(
{\bf t}_2
{\bf t}_2
)
}
}
A_1t_{2q}H_{sp}
(
{\bf t}_1
)
+
\fr1h
\fr
{
(
{\bf t}_1
{\bf t}_2
)
}
{
(
{\bf t}_1
{\bf t}_1
)
\sqrt{
(
{\bf t}_1
{\bf t}_1
)
}
\sqrt{
(
{\bf t}_2
{\bf t}_2
)
}
}
\fr1
{
u
(
{\bf t}_1,
{\bf t}_2
)
}
A_2d_{1s}t_{1p}t_{2q}
$$
\ses
\ses
$$
+
\fr1h
\fr
1
{
\sqrt{
(
{\bf t}_1
{\bf t}_1
)
}
\sqrt{
(
{\bf t}_2
{\bf t}_2
)
}
}
\fr1
{
u
(
{\bf t}_1,
{\bf t}_2
)
}
A_2
\times
$$
\ses
\ses
$$
\Bigl[
\fr1
{
(
{\bf t}_1
{\bf t}_1
)
}
\lf(
(
{\bf t}_1
{\bf t}_1
)
d_{2p}d_{2q}
+
(
{\bf t}_2
{\bf t}_2
)
d_{1p}d_{1q}
\rg)
d_{1s}
+
(
{\bf t}_1
{\bf t}_2
)
H_{ps}
(
{\bf t}_1
)
d_{2q}
-
(
{\bf t}_2
{\bf t}_2
)
H_{qs}
(
{\bf t}_1
)
d_{1p}
\Bigl]
$$
\ses
\ses
\be
+
\fr1h
\fr1
{
(
{\bf t}_1
{\bf t}_1
)
\sqrt{
(
{\bf t}_1
{\bf t}_1
)
}
\sqrt{
(
{\bf t}_2
{\bf t}_2
)
}
}
\Bigl[
(1-\fr1{h^2})\sin\al-
\fr
{
(
{\bf t}_1
{\bf t}_2
)
}
{u
(
{\bf t}_1,
{\bf t}_2
)
}
A_2
\Bigr]
d_{1s}d_{1p}d_{2q},
\ee
where
\ses\\
\be
u
(
{\bf t}_1,
{\bf t}_2
)
=
\sqrt{
(
{\bf t}_1
{\bf t}_1
)
(
{\bf t}_2
{\bf t}_2
)
-
(
{\bf t}_1
{\bf t}_2
)^2
}
\ee
\ses\\
and we used the relations:
\be
\D{\al}
{t_1^s}
=
-\fr1h
\fr
{
{\bf d}_1
}
{
(
{\bf t}_1
{\bf t}_1
)
},
\qquad
\D{\fr1u}
{t_1^s}
=
-\fr1{u^2}
{\bf d}_2,
\ee
\ses
\ses
\be
\D{A_1}
{t_1^s}=
\fr1h
\fr
{
(
{\bf t}_1
{\bf t}_2
)
}
{
(
{\bf t}_1
{\bf t}_1
)
u
(
{\bf t}_1,
{\bf t}_2
)
}
A_2d_{1s},
\ee
\ses
\ses
$$
\D{A_2}
{t_1^s}=
\fr1h
\fr
{
(
{\bf t}_1
{\bf t}_2
)
}
{
(
{\bf t}_1
{\bf t}_1
)
u
(
{\bf t}_1,
{\bf t}_2
)
}
A_1
d_{1s}
-
(1-\fr1{h^2})
\fr
{
(
{\bf t}_2
{\bf t}_2
)
}
{
u
(
{\bf t}_1,
{\bf t}_2
)
}
\fr
{
\sin\al
}
{
u
(
{\bf t}_1,
{\bf t}_2
)
}
 d_{1s}
$$
\ses\ses
\be
=
\fr{d_{1s}}
{
(
{\bf t}_1
{\bf t}_1
)
}
\Bigl[
-(1-\fr1{h^2})\sin\al+
\fr
{
(
{\bf t}_1
{\bf t}_2
)
}
{u
(
{\bf t}_1,
{\bf t}_2
)
}
A_2\Bigr].
\ee
\ses
Since
\be
\lim_{
{\bf t}_2\to
{\bf t}_1
}
\Bigl\{
{A_1}
\Bigl\}=
1-\fr1{h^2},
\qquad
\lim_{
{\bf t}_2\to
{\bf t}_1
}
\Bigl\{
\fr
{A_2}
{u}
\Bigl\}=
0,
\ee
\ses
\ses
\be
\lim_{
{\bf t}_2\to
{\bf t}_1=
{\bf t}
}
\Bigl\{
\D
{
n_{pq}(g;
{\bf t}_1,
{\bf t}_2
)
}
{
 t_1^s
}
\Bigl\}=
(1-\fr1{h^2})
\fr
{t_q}
{
(
{\bf t}
{\bf t}
)
}
H_{sp}(
{\bf t}
),
\ee
\ses
and
\ses
\ses
\be
\lim_{
{\bf t}_2\to
{\bf t}_1=
{\bf t}
}
\Bigl\{
\D
{
n_{pq}(g;
{\bf t}_1,
{\bf t}_2
)
}
{
 t_2^s
}
\Bigl\}=
(1-\fr1{h^2})
\fr
{t_p}
{
(
{\bf t}
{\bf t}
)
}
H_{sq}(
{\bf t}
),
\ee
 the fundamental consequence
\ses\\
\be
\lim_{
{\bf t}_2\to
{\bf t}_1=
{\bf t}
}
\Bigl\{
\D
{
n_{pq}(g;
{\bf t}_1,
{\bf t}_2
)
}
{
 t_1^s
}
+
\D
{
n_{pq}(g;
{\bf t}_1,
{\bf t}_2
)
}
{
 t_2^s
}
\Bigl\}=
\D
{
n_{pq}(g;
{\bf t}
)
}
{
 t^s
}
\ee
is obtained.

The expansion with respect to an appropriate orthonormal frame
$
f^R_p
(g;
{\bf t}_1,
{\bf t}_2
)
$
can be found to read
\be
n_{pq}
(g;
{\bf t}_1,
{\bf t}_2
)
=
\sum_{R=1}^{N}
f^R_p
(g;
{\bf t}_1,
{\bf t}_2
)
f^R_q
(g;
{\bf t}_2,
{\bf t}_1
)
\ee
\ses
with
$$
\sqrt
{
h
\sqrt{(
{\bf t}_1
{\bf t}_1
)}
\sqrt{(
{\bf t}_2
{\bf t}_2
)}
}
f^R_p
(g;
{\bf t}_1,
{\bf t}_2
)
=
zh^R_p
$$
\ses
\ses
\ses
\ses
$$
-\fr1{
(
{\bf t}_1
{\bf t}_2
)
}
\Bigl[
z
-
\sqrt{z^2+
(
{\bf t}_1
{\bf t}_2
)
\Bigl(
h
\cos
\al
-
(
{\bf t}_1
{\bf t}_2
)
\fr{
\sin
\al
}
{
\sqrt{
(
{\bf t}_1
{\bf t}_1
)
(
{\bf t}_2
{\bf t}_2
)
-
(
{\bf t}_1
{\bf t}_2
)^2
}
}
\Bigr)
}
\Bigr]
 t_2^R
 t_{1p}
$$
\ses
\ses
\ses
\ses
\ses
\be
+\fr1{
(
{\bf t}_1
{\bf t}_2
)}
\Bigl[
z-\sqrt{z^2+
(
{\bf t}_1
{\bf t}_2
)
\Bigl(
\fr1h
\cos
\al
-
(
{\bf t}_1
{\bf t}_2
)
\fr{
\sin
\al
}
{
\sqrt{
(
{\bf t}_1
{\bf t}_1
)
(
{\bf t}_2
{\bf t}_2
)
-
(
{\bf t}_1
{\bf t}_2
)^2
}
}
\Bigr)
}
\Bigr]
 d_{2p}
d_1^R,
\ee
where
\ses\\
\be
z=
\sqrt
{
(
{\bf t}_1
{\bf t}_1
)
(
{\bf t}_2
{\bf t}_2
)
\fr{
\sin
\al
}
{
\sqrt{
(
{\bf t}_1
{\bf t}_1
)
(
{\bf t}_2
{\bf t}_2
)
-
(
{\bf t}_1
{\bf t}_2
)^2
}
}
},
\ee
\ses
\ses
or
finally,
\ses\\
$$
\sqrt
{
h
\sqrt{(
{\bf t}_1
{\bf t}_1
)}
\sqrt{(
{\bf t}_2
{\bf t}_2
)}
}
f^R_p
(g;
{\bf t}_1,
{\bf t}_2
)
$$
\ses
\ses
\ses
\ses
$$
=
zh^R_p
-\fr1{
(
{\bf t}_1
{\bf t}_2
)
}
\Bigl[
z-\sqrt{
h
(
{\bf t}_1
{\bf t}_2
)
\cos
\al
+
\sqrt{
(
{\bf t}_1
{\bf t}_1
)
(
{\bf t}_2
{\bf t}_2
)
-
(
{\bf t}_1
{\bf t}_2
)^2
}
\sin
\al
}
\Bigr]
\,
 t_2^R
 t_{1p}
$$
\ses
\ses
\ses
\ses
\ses
\ses
\ses
\ses
\be
+\fr1{
(
{\bf t}_1
{\bf t}_2
)}
\Bigl[
z-\sqrt{
\fr1h
(
{\bf t}_1
{\bf t}_2
)
\cos
\al
+
\sqrt{
(
{\bf t}_1
{\bf t}_1
)
(
{\bf t}_2
{\bf t}_2
)
-
(
{\bf t}_1
{\bf t}_2
)^2
}
\sin
\al
}
\Bigr]
\,
 d_{2p}
d_1^R.
\ee

Contracting the frame by vectors yields
\ses\\
$$
f^R_p
(g;
{\bf t}_1,
{\bf t}_2
)
t_1^p
$$
\ses
\ses
$$
=
\fr
1
{
\sqrt
{
h
\sqrt{
(
{\bf t}_1
{\bf t}_1
)}
\sqrt{
(
{\bf t}_1
{\bf t}_2
)}
}
}
\Bigl
[
\fr
{
(
{\bf t}_1
{\bf t}_1
)
}
{
(
{\bf t}_1
{\bf t}_2
)
}
\lf(
\sqrt
{h
(
{\bf t}_1
{\bf t}_2
)
\cos\al
+
\sqrt{
(
{\bf t}_1
{\bf t}_1
)
(
{\bf t}_2
{\bf t}_2
)
-
(
{\bf t}_1
{\bf t}_2
)^2
}
\,
\sin\al
}
\rg.
$$
\ses
\ses
\ses
\ses
$$
\lf.
-
\sqrt
{\fr1h
(
{\bf t}_1
{\bf t}_2
)
\cos\al
+
\sqrt{
(
{\bf t}_1
{\bf t}_1
)
(
{\bf t}_2
{\bf t}_2
)
-
(
{\bf t}_1
{\bf t}_2
)^2
}
\,
\sin\al
}
\rg)
t_2^R
$$
\ses
\ses
\be
+
\sqrt
{\fr1h
(
{\bf t}_1
{\bf t}_2
)
\cos\al
+
\sqrt{
(
{\bf t}_1
{\bf t}_1
)
(
{\bf t}_2
{\bf t}_2
)
-
(
{\bf t}_1
{\bf t}_2
)^2
}
\,
\sin\al
}
\,
t_1^R
\Bigr],
\ee
\ses
\ses\\
$$
f^R_p
(g;
{\bf t}_1,
{\bf t}_2
)
t_2^p
$$
\ses
\ses
\be
=
\fr
1
{
\sqrt
{
h
\sqrt{
(
{\bf t}_1
{\bf t}_1
)}
\sqrt{
(
{\bf t}_2
{\bf t}_2
)}
}
}
\sqrt
{h
(
{\bf t}_1
{\bf t}_2
)
\cos\al
+
\sqrt{
(
{\bf t}_1
{\bf t}_1
)
(
{\bf t}_2
{\bf t}_2
)
-
(
{\bf t}_1
{\bf t}_2
)^2
}
\,
\sin\al
}
\,
t_2^R,
\ee
and
\ses\\
$$
\sum_{R=1}^{N}
f^R_p
(g;
{\bf t}_1,
{\bf t}_2
)
t_1^R
=
$$
\ses
\ses
\be
=
\fr
1
{
\sqrt
{
h
\sqrt{
(
{\bf t}_1
{\bf t}_1
)}
\sqrt{
(
{\bf t}_2
{\bf t}_2
)}
}
}
\sqrt
{h
(
{\bf t}_1
{\bf t}_2
)
\cos\al
+
\sqrt{
(
{\bf t}_1
{\bf t}_1
)
(
{\bf t}_2
{\bf t}_2
)
-
(
{\bf t}_1
{\bf t}_2
)^2
}
\,
\sin\al
}
\,
\,
t_{1p},
\ee
\ses\\
together with
\ses\\
$$
\sum_{R=1}^{N}
f^R_p
(g;
{\bf t}_1,
{\bf t}_2
)
t_2^R
=
$$
\ses
\ses
$$
=
\fr
1
{
\sqrt
{
h
\sqrt{
(
{\bf t}_1
{\bf t}_1
)}
\sqrt{
(
{\bf t}_1
{\bf t}_2
)}
}
}
\Bigl
[
\fr
{
(
{\bf t}_2
{\bf t}_2
)
}
{
(
{\bf t}_1
{\bf t}_2
)
}
\lf(
\sqrt
{h
(
{\bf t}_1
{\bf t}_2
)
\cos\al
+
\sqrt{
(
{\bf t}_1
{\bf t}_1
)
(
{\bf t}_2
{\bf t}_2
)
-
(
{\bf t}_1
{\bf t}_2
)^2
}
\,
\sin\al
}
\rg.
$$
\ses
\ses
\ses
\ses
$$
\lf.
-
\sqrt
{\fr1h
(
{\bf t}_1
{\bf t}_2
)
\cos\al
+
\sqrt{
(
{\bf t}_1
{\bf t}_1
)
(
{\bf t}_2
{\bf t}_2
)
-
(
{\bf t}_1
{\bf t}_2
)^2
}
\,
\sin\al
}
\rg)
t_{1p}
$$
\ses
\ses
\be
+
\sqrt
{\fr1h
(
{\bf t}_1
{\bf t}_2
)
\cos\al
+
\sqrt{
(
{\bf t}_1
{\bf t}_1
)
(
{\bf t}_2
{\bf t}_2
)
-
(
{\bf t}_1
{\bf t}_2
)^2
}
\,
\sin\al
}
\,
\,t_{2p}
\Bigr].
\ee

\ses
\ses

\setcounter{sctn}{3}
\setcounter{equation}{0}
{\nin\bf 1.3.
Covariant version
}

\ses
\ses

It proves possible to convert the approach into {\it the co-version}
 by introducing
{\it the co-vectors}
\be
 T_{1p}
(g;
{\bf t}_1,
{\bf t}_2
)
\eqdef
n_{pq}
(g;
{\bf t}_1,
{\bf t}_2
)
{\bf t}_2^q,
\qquad
 T_{2q}
(g;
{\bf t}_1,
{\bf t}_2
)
\eqdef
{\bf t}_1^p
n_{pq}(g;
{\bf t}_1,
{\bf t}_2
).
\ee
Using (2.2),
we get
\ses\\
\be
{\bf T}_1
=
\fr
{
\sqrt
{
(
{\bf t}_2
{\bf t}_2
)
}
}
{
\sqrt
{
(
{\bf t}_1
{\bf t}_1
)
}
}
{\bf t}_1
\,
\cos
\al
+
\fr{
\sqrt{(
{\bf t}_2
{\bf t}_2
)}
}
{
h
\sqrt{(
{\bf t}_1
{\bf t}_1
)}
}
{\bf d}_1
\sin
\al
\ee
and
\be
{\bf T}_2
=
\fr
{
\sqrt
{
(
{\bf t}_1
{\bf t}_1
)
}
}
{
\sqrt
{
(
{\bf t}_2
{\bf t}_2
)
}
}
{\bf t}_2
\,
\cos
\al
+
\fr{
\sqrt{(
{\bf t}_1
{\bf t}_1
)}
}
{
h
\sqrt{(
{\bf t}_2
{\bf t}_2
)}
}
{\bf d}_2
\sin
\al.
\ee
The equality
\ses\\
\be
{\bf t}_1
{\bf T}_1
+
{\bf t}_2
{\bf T}_2
=
2
<{\bf t}_1,{\bf t}_2>
=
2
\sqrt{(
{\bf t}_1
{\bf t}_1
)}
\sqrt{(
{\bf t}_2
{\bf t}_2
)}
\cos\al
\ee
holds.
Also,
\be
\lim_{
{\bf t}_2\to
{\bf t}_1=
{\bf t}
}
\Bigl\{
{\bf T}_1
\Bigl\}=
\lim_{
{\bf t}_2\to
{\bf t}_1=
{\bf t}
}
\Bigl\{
{\bf T}_2
\Bigl\}=
{\bf t}.
\ee
The metric tensor (2.1)-(2.2) is obtainable from these vectors as follows:
\be
n_{pq}(g;
{\bf t}_1,
{\bf t}_2
)
=
\D
{T_{1p}}
{t_2^q}
=
\D
{T_{2q}}
{t_1^p}.
\ee

\ses

The respective products are found to be
\be
(
{\bf T}_1
{\bf T}_1
)
=
(
{\bf t}_2
{\bf t}_2
)
(\cos^2\al+\fr1{h^2}\sin^2\al),
\qquad
(
{\bf T}_2
{\bf T}_2
)
=
(
{\bf t}_1
{\bf t}_1
)
(\cos^2\al+\fr1{h^2}\sin^2\al),
\ee
and
\ses\\
\be
(
{\bf T}_1
{\bf T}_2
)
=
(\cos^2\al-\fr1{h^2}\sin^2\al)
(
{\bf t}_1
{\bf t}_2
)
+
2\fr1h
\sqrt{
(
{\bf t}_1
{\bf t}_1
)
(
{\bf t}_2
{\bf t}_2
)
-
(
{\bf t}_1
{\bf t}_2
)^2
}
\,
\cos
\al
\sin\al,
\ee
\ses
\ses\\
together with
$$
\sqrt
{
(
{\bf T}_1
{\bf T}_1
)
(
{\bf T}_2
{\bf T}_2
)
-
(
{\bf T}_1
{\bf T}_2
)^2
}
$$
\ses
\ses
\be
=
\fr2h
(
{\bf t}_1
{\bf t}_2
)
\sin\al\cos\al
-
(\cos^2\al-\fr1{h^2}\sin^2\al)
\sqrt{
(
{\bf t}_1
{\bf t}_1
)
(
{\bf t}_2
{\bf t}_2
)
-
(
{\bf t}_1
{\bf t}_2
)^2
}.
\ee

From  (3.7)-(3.8) it follows that
\ses\\
$$
(\cos^2\al+\fr1{h^2}\sin^2\al)^2
\sqrt{
(
{\bf t}_1
{\bf t}_1
)
(
{\bf t}_2
{\bf t}_2
)
-
(
{\bf t}_1
{\bf t}_2
)^2
}
$$
\ses
\ses
\be
=
\fr2h
(
{\bf T}_1
{\bf T}_2
)
\sin\al\cos\al
-
(\cos^2\al-\fr1{h^2}\sin^2\al)
\sqrt
{
(
{\bf T}_1
{\bf T}_1
)
(
{\bf T}_2
{\bf T}_2
)
-
(
{\bf T}_1
{\bf T}_2
)^2
}
\ee
and
\ses\\
$$
(\cos^2\al+\fr1{h^2}\sin^2\al)^2
(
{\bf t}_1
{\bf t}_2
)
$$
\ses
\be
=
(\cos^2\al-\fr1{h^2}\sin^2\al)
(
{\bf T}_1
{\bf T}_2
)
+
\fr2h
\sin\al\cos\al
\sqrt
{
(
{\bf T}_1
{\bf T}_1
)
(
{\bf T}_2
{\bf T}_2
)
-
(
{\bf T}_1
{\bf T}_2
)^2
},
\ee
\ses
\ses
together with
\ses\\
$$
(\cos^2\al+\fr1{h^2}\sin^2\al)
\Bigl[
-\fr1h
(
{\bf t}_1
{\bf t}_2
)
\sin\al
+
\sqrt{
(
{\bf t}_1
{\bf t}_1
)
(
{\bf t}_2
{\bf t}_2
)
-
(
{\bf t}_1
{\bf t}_2
)^2
}
\,
\cos\al
\Bigr]
$$
\ses
\ses
\be
=
\fr1h
(
{\bf T}_1
{\bf T}_2
)
\sin\al
-
\sqrt
{
(
{\bf T}_1
{\bf T}_1
)
(
{\bf T}_2
{\bf T}_2
)
-
(
{\bf T}_1
{\bf T}_2
)^2
}
\,
\cos\al.
\ee

Using these formulas in calculating the co-representation
\be
\al=\al(h;
{\bf T}_1,
{\bf T}_2
)
\ee
for the angle (1.18) yields the following implicit equation:
\be
\cos(h\al)=
\fr
{
(\cos^2\al-\fr1{h^2}\sin^2\al)
(
{\bf T}_1
{\bf T}_2
)
+
\fr2h
\sin\al\cos\al
\sqrt
{
(
{\bf T}_1
{\bf T}_1
)
(
{\bf T}_2
{\bf T}_2
)
-
(
{\bf T}_1
{\bf T}_2
)^2
}
}
{
(\cos^2\al+\fr1{h^2}\sin^2\al)
\sqrt{
(
{\bf T}_1
{\bf T}_1
)
}
\sqrt{
(
{\bf T}_2
{\bf T}_2
)
}
}.
\ee

The respective co-version of the scalar product (1.30) reads
\be
<{\bf T}_2,{\bf T}_2>
=
\sqrt{
(
{\bf T}_1
{\bf T}_1
)
}
\sqrt{
(
{\bf T}_2
{\bf T}_2
)
}
\cos\al.
\ee

On this way the set (3.2)-(3.3) can be inverted, yielding
\ses\\
$$
{\bf t}_1
(g;
{\bf T}_1,
{\bf T}_2
)
=
\fr1f
\Bigl[
\fr
{
\sqrt{
(
{\bf t}_1
{\bf t}_1
)
}
}{
\sqrt{
(
{\bf t}_2
{\bf t}_2
)
}
}
\lf(
\cos\al-\fr1h
\fr
{
(
{\bf t}_1
{\bf t}_2
)
}
{
\sqrt{
(
{\bf t}_1
{\bf t}_1
)
(
{\bf t}_2
{\bf t}_2
)
-
(
{\bf t}_1
{\bf t}_2
)^2
}
}
\sin\al
\rg)
{\bf T}_1
$$
\ses
\ses
\be
-
\fr1h
\fr
{
\sqrt{
(
{\bf t}_1
{\bf t}_1
)
}
\sqrt{
(
{\bf t}_2
{\bf t}_2
)
}
}
{
\sqrt{
(
{\bf t}_1
{\bf t}_1
)
(
{\bf t}_2
{\bf t}_2
)
-
(
{\bf t}_1
{\bf t}_2
)^2
}
}
\sin\al
\,\,
{\bf T}_2
\Bigl]
\ee
and
\ses
\ses\\
$$
{\bf t}_2
(g;
{\bf T}_1,
{\bf T}_2
)
=
\fr1f
\Bigl[
\fr
{
\sqrt{
(
{\bf t}_2
{\bf t}_2
)
}
}{
\sqrt{
(
{\bf t}_1
{\bf t}_1
)
}
}
\lf(
\cos\al-\fr1h
\fr
{
(
{\bf t}_1
{\bf t}_2
)
}
{
\sqrt{
(
{\bf t}_1
{\bf t}_1
)
(
{\bf t}_2
{\bf t}_2
)
-
(
{\bf t}_1
{\bf t}_2
)^2
}
}
\sin\al
\rg)
{\bf T}_2
$$
\ses
\ses
\be
-
\fr1h
\fr
{
\sqrt{
(
{\bf t}_1
{\bf t}_1
)
}
\sqrt{
(
{\bf t}_2
{\bf t}_2
)
}
}
{
\sqrt{
(
{\bf t}_1
{\bf t}_1
)
(
{\bf t}_2
{\bf t}_2
)
-
(
{\bf t}_1
{\bf t}_2
)^2
}
}
\sin\al
\,\,
{\bf T}_1
\Bigl],
\ee
where
\ses\\
$$
f=
\lf(
\cos\al-\fr1h
\fr
{
(
{\bf t}_1
{\bf t}_2
)
}
{
\sqrt{
(
{\bf t}_1
{\bf t}_1
)
(
{\bf t}_2
{\bf t}_2
)
-
(
{\bf t}_1
{\bf t}_2
)^2
}
}
\sin\al
\rg)^2
$$
\ses
\ses
\be
-
\fr1{h^2}
\lf
(
\fr
{
\sin\al
}
{
\sqrt{
(
{\bf t}_1
{\bf t}_1
)
(
{\bf t}_2
{\bf t}_2
)
-
(
{\bf t}_1
{\bf t}_2
)^2
}
}
\rg)^2
(
{\bf t}_1
{\bf t}_1
)
(
{\bf t}_2
{\bf t}_2
),
\ee
or
\ses\\
\be
f=
\cos^2\al-\fr1{h^2}\sin^2\al
-\fr2h
\fr
{
\sin\al
\cos\al
}
{
\sqrt{
(
{\bf t}_1
{\bf t}_1
)
(
{\bf t}_2
{\bf t}_2
)
-
(
{\bf t}_1
{\bf t}_2
)^2
}
}
(
{\bf t}_1
{\bf t}_2
).
\ee
Taking into account (3.9), this function can be written merely as
\ses\\
\be
f=
-
\fr{
\sqrt{
(
{\bf T}_1
{\bf T}_1
)
(
{\bf T}_2
{\bf T}_2
)
-
(
{\bf T}_1
{\bf T}_2
)^2
}
}
{
\sqrt{
(
{\bf t}_1
{\bf t}_1
)
(
{\bf t}_2
{\bf t}_2
)
-
(
{\bf t}_1
{\bf t}_2
)^2
}
}.
\ee

Thus we find
\ses\\
\be
{\bf t}_1
=
\fr
{
\sqrt
{
(
{\bf T}_2
{\bf T}_2
)
}
}
{
\sqrt
{
(
{\bf T}_1
{\bf T}_1
)
}
}
{\bf T}_1
\,
\fr{\cos\al}
{\cos^2\al+\fr1{h^2}\sin^2\al}
+
\fr1h
\fr
{
\sqrt
{
(
{\bf T}_2
{\bf T}_2
)
}
}
{
\sqrt
{
(
{\bf T}_1
{\bf T}_1
)
}
}
{\bf D}_1
\fr{\sin\al}
{\cos^2\al+\fr1{h^2}\sin^2\al}
\ee
and
\be
{\bf t}_2
=
\fr
{
\sqrt
{
(
{\bf T}_1
{\bf T}_1
)
}
}
{
\sqrt
{
(
{\bf T}_2
{\bf T}_2
)
}
}
{\bf T}_2
\,
\fr{\cos\al}
{\cos^2\al+\fr1{h^2}\sin^2\al}
+
\fr1h
\fr
{
\sqrt
{
(
{\bf T}_1
{\bf T}_1
)
}
}
{
\sqrt
{
(
{\bf T}_2
{\bf T}_2
)
}
}
{\bf D}_2
\fr{\sin\al}
{\cos^2\al+\fr1{h^2}\sin^2\al},
\ee
where
\ses\\
\be
{\bf D}_1
=
\fr{
(
{\bf T}_1
{\bf T}_1
)
{\bf T}_2
-
(
{\bf T}_1
{\bf T}_2
)
{\bf T}_1
}
{
\sqrt
{(
{\bf T}_1
{\bf T}_1
)
(
{\bf T}_2
{\bf T}_2
)
-
(
{\bf T}_1
{\bf T}_2
)^2
}
},
\qquad
{\bf D}_2
=
\fr{
(
{\bf T}_2
{\bf T}_2
)
{\bf T}_1
-
(
{\bf T}_1
{\bf T}_2
)
{\bf T}_2
}
{
\sqrt
{
(
{\bf T}_1
{\bf T}_1
)
(
{\bf T}_2
{\bf T}_2
)
-
(
{\bf T}_1
{\bf T}_2
)^2
}
}.
\ee

\ses
\ses
\ses

Similarly to (1.38)-(1.40), the identities
\be
(
{\bf T}_1
{\bf D}_1
)=0,
\qquad
(
{\bf T}_2
{\bf D}_2
)=0,
\ee
\ses
\be
(
{\bf D}_1
{\bf D}_2
)=-
(
{\bf T}_1
{\bf T}_2
),
\qquad
(
{\bf D}_1
{\bf D}_1
)=
(
{\bf T}_1
{\bf T}_1
),
\qquad
(
{\bf D}_2
{\bf D}_2
)=
(
{\bf T}_2
{\bf T}_2
),
\ee
and
\ses\\
\be
(
{\bf D}_1
{\bf T}_2
)
=
(
{\bf T}_1
{\bf D}_2
)
=
\sqrt
{
(
{\bf T}_1
{\bf T}_1
)
(
{\bf T}_2
{\bf T}_2
)
-
(
{\bf T}_1
{\bf T}_2
)^2
}
\ee
hold.

By the help of (3.20)-(3.21), and in close similarity to (3.6),
the co-version
\be
N^{pq}(g;
{\bf T}_1,
{\bf T}_2
)
\eqdef
\D
{t_1^p}
{T_{2q}}
=
\D
{t_2^q}
{T_{1p}}
\ee
of the two-vector metric tensor (2.2) can be arrived at.

\ses
\ses

\setcounter{sctn}{4}
\setcounter{equation}{0}
{\nin\bf 1.4.
$\cE_g^{PD}$-parallelogram law
}

\ses
\ses

Let
$
{\bf t}_1,
 {\bf t}_2$, and
$ {\bf t}_3 $ be three vectors issued from the same origin $O$,
subject to the conditions that the angle between $ {\bf t}_1 $ and
$ {\bf t}_2 $ is acute and the vector $ {\bf t}_3 $ is positioned
between the vectors $ {\bf t}_1 $ and $ {\bf t}_2 $. Let us denote
the end points of the vectors $ {\bf t}_1,
 {\bf t}_2$, and
$
{\bf t}_3
$
as
$X_1, X_2$, and $X_3$,
respectively.
On joining the points
$X_1$ and $X_3$, and also
$X_2$ and $X_3$,
by  $\cE_g^{PD}$-geodesics, we get a tetragonal figure, to be denoted as
$\cP_4$.

Using Eqs. (1.14), (1.15), and (1.27), we can set forth
the following couple of two equations:
\be
({\bf t}_2
{\bf t}_2
)
=
({\bf t}_1
{\bf t}_1
)
+
({\bf t}_3
{\bf t}_3
)
-2
\sqrt
{
({\bf t}_1
{\bf t}_1
)
}
\,
\sqrt
{
({\bf t}_3
{\bf t}_3
)
}
\cos
\Bigl[
\fr1h\arccos
\fr{
(
{\bf t}_1
{\bf t}_3
)}
{
\sqrt{(
{\bf t}_1
{\bf t}_1
)}
\,
\sqrt{(
{\bf t}_3
{\bf t}_3
)}
}
\Bigr]
\ee
and
\ses\\
\be
({\bf t}_1
{\bf t}_1
)
=
({\bf t}_3
{\bf t}_3
)
+
({\bf t}_2
{\bf t}_2
)
-2
\sqrt
{
({\bf t}_3
{\bf t}_3
)
}
\,
\sqrt
{
({\bf t}_2
{\bf t}_2
)
}
\cos
\Bigl[
\fr1h\arccos
\fr{
(
{\bf t}_2
{\bf t}_3
)}
{
\sqrt{(
{\bf t}_2
{\bf t}_2
)}
\,
\sqrt{(
{\bf t}_3
{\bf t}_3
)}
}
\Bigr],
\ee
\ses\\
which can also be rewritten in the convenient form
\be
\sqrt{(
{\bf t}_3
{\bf t}_3
)}
-
\fr{
({\bf t}_2
{\bf t}_2
)
-
({\bf t}_1
{\bf t}_1
)
}
{
\sqrt{(
{\bf t}_3
{\bf t}_3
)}
}
=2
\sqrt
{
({\bf t}_1
{\bf t}_1
)
}
\cos
\Bigl[
\fr1h\arccos
\fr{
(
{\bf t}_1
{\bf t}_3
)}
{
\sqrt{(
{\bf t}_1
{\bf t}_1
)}
\,
\sqrt{(
{\bf t}_3
{\bf t}_3
)}
}
\Bigr]
\ee
and
\ses\\
\be
\sqrt{(
{\bf t}_3
{\bf t}_3
)}
-
\fr{
({\bf t}_1
{\bf t}_1
)
-
({\bf t}_2
{\bf t}_2
)
}
{
\sqrt{(
{\bf t}_3
{\bf t}_3
)}
}
=2
\sqrt
{
({\bf t}_2
{\bf t}_2
)
}
\cos
\Bigl[
\fr1h\arccos
\fr{
(
{\bf t}_2
{\bf t}_3
)}
{
\sqrt{(
{\bf t}_2
{\bf t}_2
)}
\,
\sqrt{(
{\bf t}_3
{\bf t}_3
)}
}
\Bigr].
\ee

In (4.1), the left-hand part is the squared length of the straight side $OX_2$
and the
right-hand side is the squared length of the geodesic side
$X_1X_3$. According to (4.2),
the lengths of $OX_1$
and $X_2X_3$ are equal.
Under these conditions, the figure $\cP_4$ does attribute the general
property of the euclidean parallelogram that the lengths of opposite sides
are equal. In this vein, we introduce the following

\ses

{\bf DEFINITION}. Subject to the equations (4.1) and (4.2),
the tetragonal figure
$\cP_4$
is called
{
\it
the
$\cE_g^{PD}$-parallelogram,}
and the vector
$
{\bf t}_3
$
is called
the
$\cE_g^{PD}$-sum vector:
\be
{\bf t}_3
=
{\bf t}_1
\oplus
{\bf t}_2.
\ee

\ses

{\bf NOTE}. The qualitative distinction here
from euclidean patterns
is that the  sides
$X_1X_3$ and $X_3X_2$ of the $\cP_4$ are curved lines in general,
namely  geodesic arcs,
which generally cease to be straight under the $\cE_g^{PD}$-extension.

Finding the sum vector (4.5)
implies solving the set of the equations (4.3) and (4.4).
It proves easy to proceed  approximately, namely taking
\be
\fr1h=1+k
\ee
and
\be
{\bf t}_3
=
{\bf t}_1
+{\bf t}_2
+k
{\bf c}
({\bf t}_1,{\bf t}_2),
\quad
k\ll 1.
\ee
Under these conditions, on inserting (4.6) and (4.7) in (4.3), we find
\ses\\
$$
\sqrt
{
(
{\bf t}_2
+{\bf t}_1
)^2
}
+
k\fr{
(
{\bf t}_2
+{\bf t}_1
){\bf c}
}
{
\sqrt
{
(
{\bf t}_2
+{\bf t}_1
)^2
}
}
-
\fr{
({\bf t}_2
{\bf t}_2
)
-
({\bf t}_1
{\bf t}_1
)
}
{
\sqrt
{
(
{\bf t}_2
+{\bf t}_1
)^2
}
}
\lf(1-k\fr{
(
{\bf t}_2
+{\bf t}_1
){\bf c}
}
{
{
(
{\bf t}_2
+{\bf t}_1
)^2
}
}
\rg)
$$
\ses
\ses
$$
=2
\sqrt
{
({\bf t}_1
{\bf t}_1
)
}
\cos
\Bigl[
(1+k)\arccos
\fr{
(
{\bf t}_1
{\bf t}_3
)}
{
\sqrt{(
{\bf t}_1
{\bf t}_1
)}
\,
\sqrt{(
{\bf t}_3
{\bf t}_3
)}
}
\Bigr]
$$
\ses
\ses
\ses
\ses
$$
=2
\fr{
(
{\bf t}_1
{\bf t}_3
)}
{
\sqrt{(
{\bf t}_3
{\bf t}_3
)}
}
-
2
k
\sqrt
{
({\bf t}_1
{\bf t}_1
)
}
\sqrt
{
1-
\lf(
\fr{
{\bf t}_1
(
{\bf t}_2
+
{\bf t}_1
)
}
{
\sqrt{(
{\bf t}_1
{\bf t}_1
)}
\,
\sqrt
{
(
{\bf t}_2
+{\bf t}_1
)^2
}
}
\rg)^2
}
\arccos
\fr{
{\bf t}_1
(
{\bf t}_2
+
{\bf t}_1
)
}
{
\sqrt{(
{\bf t}_1
{\bf t}_1
)}
\,
\sqrt
{
(
{\bf t}_2
+{\bf t}_1
)^2
}
}
$$
\ses
\ses
\ses
\ses
\ses
$$
=
2k
\fr{
(
{\bf t}_1
{\bf c}
)}
{
\sqrt
{
(
{\bf t}_2
+{\bf t}_1
)^2
}
}
+
2
\fr{
{\bf t}_1
(
{\bf t}_2
+
{\bf t}_1
)}
{
\sqrt
{
(
{\bf t}_2
+{\bf t}_1
)^2
}
}
\lf(1-k
\fr{
(
{\bf t}_2
+{\bf t}_1
){\bf c}
}
{
{
(
{\bf t}_2
+{\bf t}_1
)^2
}
}
\rg)
$$
\ses
\ses
\be
-
2k
\sqrt
{
({\bf t}_1
{\bf t}_1
)
}
\sqrt
{
1-
\lf(
\fr{
{\bf t}_1
(
{\bf t}_2
+
{\bf t}_1
)
}
{
\sqrt{(
{\bf t}_1
{\bf t}_1
)}
\,
\sqrt
{
(
{\bf t}_2
+{\bf t}_1
)^2
}
}
\rg)^2
}
\arccos
\fr{
{\bf t}_1
(
{\bf t}_2
+
{\bf t}_1
)
}
{
\sqrt{(
{\bf t}_1
{\bf t}_1
)}
\,
\sqrt
{
(
{\bf t}_2
+{\bf t}_1
)^2
}
},
\ee
\ses
which entails
$$
\fr{
(
{\bf t}_2
+{\bf t}_1
){\bf c}
}
{
\sqrt
{
(
{\bf t}_2
+{\bf t}_1
)^2
}
}
+
\fr{
({\bf t}_2
{\bf t}_2
)
-
({\bf t}_1
{\bf t}_1
)
}
{
{
(
{\bf t}_2
+{\bf t}_1
)^2
}
}
\fr{
(
{\bf t}_2
+{\bf t}_1
){\bf c}
}
{
\sqrt
{
(
{\bf t}_2
+{\bf t}_1
)^2
}
}
$$
\ses
\ses
\ses
\ses
\ses
$$
=
2
\fr{
(
{\bf t}_1
{\bf c}
)}
{
\sqrt
{
(
{\bf t}_2
+{\bf t}_1
)^2
}
}
-
2
\fr{
{\bf t}_1
(
{\bf t}_2
+
{\bf t}_1
)}
{
{
(
{\bf t}_2
+{\bf t}_1
)^2
}
}
\fr{
(
{\bf t}_2
+{\bf t}_1
){\bf c}
}
{
\sqrt
{
(
{\bf t}_2
+{\bf t}_1
)^2
}
}
$$
\ses
\ses
\be
-
2
\sqrt
{
({\bf t}_1
{\bf t}_1
)
}
\sqrt
{
1-
\lf(
\fr{
{\bf t}_1
(
{\bf t}_2
+
{\bf t}_1
)
}
{
\sqrt{(
{\bf t}_1
{\bf t}_1
)}
\,
\sqrt
{
(
{\bf t}_2
+{\bf t}_1
)^2
}
}
\rg)^2
}
\arccos
\fr{
{\bf t}_1
(
{\bf t}_2
+
{\bf t}_1
)
}
{
\sqrt{(
{\bf t}_1
{\bf t}_1
)}
\,
\sqrt
{
(
{\bf t}_2
+{\bf t}_1
)^2
}
}.
\ee

Therefore we obtain
\be
{\bf t}_2
{\bf c}
=
-
u
(
{\bf t}_1,
{\bf t}_2
)
\arccos
\fr{
{\bf t}_1
(
{\bf t}_2
+
{\bf t}_1
)
}
{
\sqrt{(
{\bf t}_1
{\bf t}_1
)}
\,
\sqrt
{
(
{\bf t}_2
+{\bf t}_1
)^2
}
}.
\ee
\ses
Similarly,
from (4.4)
it follows that
\ses\\
\be
{\bf t}_1
{\bf c}
=
-
u
(
{\bf t}_1,
{\bf t}_2
)
\arccos
\fr{
{\bf t}_2
(
{\bf t}_2
+
{\bf t}_1
)
}
{
\sqrt{(
{\bf t}_2
{\bf t}_2
)}
\,
\sqrt
{
(
{\bf t}_2
+
{\bf t}_1
)^2
}
},
\ee
where
$
u
(
{\bf t}_1,
{\bf t}_2
)
$
is the function (2.8).

If we use now the symmetrized expansion
\be
{\bf c} =
m
(
{\bf t}_1,
{\bf t}_2
)
{\bf t}_1
+
n
(
{\bf t}_1 ,
{\bf t}_2
)
{\bf t}_2,
\ee
then we find
\ses\\
$$
m
(
{\bf t}_1,
{\bf t}_2
)
=
$$
\ses
\ses
\be
\fr1
{
u
(
{\bf t}_1,
{\bf t}_2
)
}
\lf(
(
{\bf t}_1
{\bf t}_2
)
\arccos
\fr{
{\bf t}_1
(
{\bf t}_2
+
{\bf t}_1
)
}
{
\sqrt{(
{\bf t}_1
{\bf t}_1
)}
\,
\sqrt
{
(
{\bf t}_2
+{\bf t}_1
)^2
}
}
-
(
{\bf t}_2
{\bf t}_2
)
\arccos
\fr{
{\bf t}_2
(
{\bf t}_2
+
{\bf t}_1
)
}
{
\sqrt{(
{\bf t}_2
{\bf t}_2
)}
\,
\sqrt
{
(
{\bf t}_2
+{\bf t}_1
)^2
}
}
\rg)
\ee
\ses
and
\ses\\
$$
n
(
{\bf t}_1,
{\bf t}_2
)
=
$$
\ses
\ses
\be
\fr1
{
u
(
{\bf t}_1,
{\bf t}_2
)
}
\lf(
(
{\bf t}_1
{\bf t}_2
)
\arccos
\fr{
{\bf t}_2
(
{\bf t}_2
+
{\bf t}_1
)
}
{
\sqrt{(
{\bf t}_2
{\bf t}_2
)}
\,
\sqrt
{
(
{\bf t}_2
+{\bf t}_1
)^2
}
}
-
(
{\bf t}_1
{\bf t}_1
)
\arccos
\fr{
{\bf t}_1
(
{\bf t}_2
+
{\bf t}_1
)
}
{
\sqrt{(
{\bf t}_1
{\bf t}_1
)}
\,
\sqrt
{
(
{\bf t}_2
+{\bf t}_1
)^2
}
}
\rg).
\ee

Since
\be
m
(
{\bf t}_1,
{\bf t}_2
)
=
n
(
{\bf t}_2,
{\bf t}_1
),
\ee
 we  just deduce the approximate solution
\ses
\be
{\bf t}_1
\oplus
{\bf t}_2
\approx
{\bf t}_1
+{\bf t}_2
+(\fr1h-1)
\Bigl
(
m
({\bf t}_1,{\bf t}_2)
{\bf t}_1
+
m
({\bf t}_2,{\bf t}_1)
{\bf t}_2
\Bigr
),
\qquad
\fr1h-1\ll 1.
\ee

Alternatively, the solution
$
{\bf t}_2
=
{\bf t}_2
({\bf t}_1,{\bf t}_3)
$
to the set of equations (4.1)-(4.2) can naturally be treated as
{
\it
the
$\cE_g^{PD}$-difference} of vectors
$
{\bf t}_3
$
and
$
{\bf t}_1
$:
\be
{\bf t}_2
=
{\bf t}_3
\ominus
{\bf t}_1.
\ee
Again, restricting ourselves to the approximation,
 from (4.1)-(4.2) we obtain
\be
{\bf t}_3
\ominus
{\bf t}_1
\approx
{\bf t}_3
-
{\bf t}_1
+
(
\fr1h-1
)
{\bf s}
(
{\bf t}_1,
{\bf t}_3
), \qquad
\fr1h-1\ll 1,
\ee
with
$$
{\bf s}
(
{\bf t}_1,
{\bf t}_3
)
=
\fr1
{
u
(
{\bf t}_1,
{\bf t}_3
)
}
\Bigl\{
\Bigl[
(
{\bf t}_1
{\bf t}_1
)
\arccos
\fr{
(
{\bf t}_1
{\bf t}_3
)
}
{
\sqrt{(
{\bf t}_1
{\bf t}_1
)}
\,
\sqrt
{
(
{\bf t}_3
{\bf t}_3
)
}
}
$$
\ses
\ses
$$
-
(
{\bf t}_3
-
{\bf t}_1,
{\bf t}_1
)
\arccos
\fr{
(
{\bf t}_3-
{\bf t}_1,
{\bf t}_3
)
}
{
\sqrt{(
{\bf t}_3 -
{\bf t}_1,
{\bf t}_3 -
{\bf t}_1
)}
\,
\sqrt
{
(
{\bf t}_3
{\bf t}_3
)
}
}
\Bigr]
(
{\bf t}_3-
{\bf t}_1
)
$$
\ses
\ses
$$
+
\Bigl[
(
{\bf t}_3-
{\bf t}_1,
{\bf t}_3 -
{\bf t}_1
)
\arccos
\fr{
(
{\bf t}_3-
{\bf t}_1,
{\bf t}_3
)
}
{
\sqrt{(
{\bf t}_3 -
{\bf t}_1,
{\bf t}_3 -
{\bf t}_1
)}
\,
\sqrt
{
(
{\bf t}_3
{\bf t}_3
)
}
}
$$
\ses
\ses
\be
-
(
{\bf t}_3
-
{\bf t}_1,
{\bf t}_1
)
\arccos
\fr{
(
{\bf t}_1
{\bf t}_3
)
}
{
\sqrt{(
{\bf t}_1
{\bf t}_1
)}
\,
\sqrt
{
(
{\bf t}_3
{\bf t}_3
)
}
}
\Bigr]
{\bf t}_1
\Bigl\}.
\ee
\ses
Here it is useful to note that
\be
(
{\bf t}_3
-
{\bf t}_1,
{\bf s}
)
= u
(
{\bf t}_1,
{\bf t}_3
)
\arccos
\fr{
(
{\bf t}_1
{\bf t}_3
)
}
{
\sqrt{(
{\bf t}_1
{\bf t}_1
)}
\,
\sqrt
{
(
{\bf t}_3
{\bf t}_3
)
}
},
\ee
\ses
\ses
\be
(
{\bf t}_1,
{\bf s}
)
=
u
(
{\bf t}_1,
{\bf t}_3
)
\arccos
\fr{
(
{\bf t}_3-
{\bf t}_1,
{\bf t}_3
)
}
{
\sqrt{(
{\bf t}_3-
{\bf t}_1,
{\bf t}_3-
{\bf t}_1
)}
\,
\sqrt
{
(
{\bf t}_3
{\bf t}_3
)
}
},
\ee
\ses\\
and
\be
u
(
{\bf t}_3-
{\bf t}_1,
{\bf t}_3
)
=
u
(
{\bf t}_1,
{\bf t}_3
).
\ee
The problem of finding
the sum vector
$
{\bf t}_1
\oplus
{\bf t}_2
$
and
the difference vector
$
{\bf t}_3
\ominus
{\bf t}_2
$
in  general exact forms
is open and seems to be difficult.

\ses
\ses

\setcounter{sctn}{5}
\setcounter{equation}{0}
{\nin\bf 1.5.
Return to initial $\cE_g^{PD}$-space
}

\ses
\ses

Applying the quasi-euclidean transformation (see (3.11) in Part II) to (1.28)
results in the following
{\it
$\cE_g^{PD}$-scalar product}:
\be
<R,S>=K(g;R)K(g;S)
\cos
\Bigl[
\fr1h\arccos\fr
{
A(g;R)A(g;S)+h^2r_{be}R^bS^e
}
{
\sqrt{B(g;R)}\,\sqrt{B(g;S)}
}
\Bigl],
\ee
\ses
\ses
\ses
so that
{\it
the $\cE_g^{PD}$-angle}
\be
\al
=
\fr1h\arccos
\fr
{
A(g;R)A(g;S)+h^2r_{be}R^bS^e
}
{
\sqrt{B(g;R)}\,\sqrt{B(g;S)}
}
\ee
is appeared;
the functions $B, K$ and $A$ can be found in Sec. 2 of Part II.

Differentiating (5.1) yields
\ses
\ses\\
\be
\D{<R,S>}{R^p}=R_p
\fr{
<R,S>
}
{K^2(g;R)}
+hK(g;S)s_p(g;R,S)\sin
\al
\ee
and
\ses
\ses\\
\be
\D{<R,S>}{S^q}=S_q
\fr{
<R,S>
}
{K^2(g;S)}
+hK(g;R)s_p(g;S,R)\sin
\al.
\ee
\ses
\ses
For the associated
{\it
$\cE_g^{PD}$-two-vector metric tensor}
\be
G_{pq}(g;R,S)
\eqdef
\Dd{<R,S>}{S^q}{R^p}
\ee
we can find explicitly the representation
\ses\
$$
G_{pq}(g;R,S)
=
\lf(
\fr{R_p}
{
K(g;R)
}
\fr
{S_q}
{K(g;S)
}
-h^2s_p(g;R,S)s_q(g;S,R)
\rg)
\cos\al
$$
\ses
\ses
\ses
\ses
\be
+
h\Bigr[
\lf(
\fr{R_p}
{
K(g;R)
}
s_q(g;S,R)
+
\fr{S_q}
{
K(g;S)
}
s_p(g;R,S)
\rg)
+s_{pq}(g;R,S)
\Bigr]
\sin\al,
\ee
\ses
\ses

where
\ses\\
$$
s_p(g;R,S)=
\fr{M_p(g;R,S)}
{
W(g;R,S)
}
\fr{K(g;R)}{B(g;R)}
$$
and
\ses\\
$$
s_{pq}(g;R,S)=
K(g;S)\D{s_p(g;R,S)}{S^q}
$$
with
\ses\\
$$
W(g;R,S)
=
\sqrt{
B(g;R)B(g;S)-
\Bigl[
A(g;R)A(g;S)+h^2r_{be}R^bS^e
\Bigl]^2
}
$$
and
\ses\\
$$
M_p(g;R,S)=B(g;R)
\sqrt{B(g;R)}\,\sqrt{B(g;S)}
\fr1{h^2}\fr{\partial}{\partial R^p}
\fr
{
A(g;R)A(g;S)+h^2r_{be}R^bS^e
}
{
\sqrt{B(g;R)}\,\sqrt{B(g;S)}
}.
$$
The latter vector  has the components
\ses
$$
h^2M_N(g;R,S)=
B(g;R)A(g;S)-
\Bigl[
A(g;R)A(g;S)+h^2r_{be}R^bS^e
\Bigl]
A(g;R)
$$
and
$$
h^2M_a(g;R,S)=B(g;R)
\lf(\fr12g
\fr{R^b}{q(R)}
A(g;S)+h^2S^b\rg)
r_{ab}
$$
\ses
\ses
$$
-
\Bigl[
A(g;R)A(g;S)+h^2r_{be}R^bS^e
\Bigl]
\lf(\fr12gR^N
+q(R)
\rg)
\fr{R^b}{q(R)}
r_{ab},
$$
\ses
\ses
\ses
\ses
which can be simplified to get
$$
M_N(g;R,S)=
q^2(R)A(g;S)-
r_{be}R^bS^e
A(g;R)
$$
and
$$
M_a(g;R,S)=
\lf(
-R^NR^b
A(g;S)+S^bB(g;R)
-
r_{ec}R^eS^c
\lf(q(R)+\fr12gR^N\rg)
\fr{R^b}{q(R)}
\rg)
r_{ab}.
$$

The identity
\be
M_p(g;R,S)R^p=0
\ee
holds.

The symmetry
\ses
\ses
\be
G_{pq}(g;
R,S)=
G_{qp}(g;
S,R)
\ee
\ses
and the Finslerian limit
\be
\lim_{
S^p\to
R^p
}
\Bigl\{
G_{pq}(g;
R,S)
\Bigl\}=
g_{pq}(g;R)
\ee
can straightforwardly be verified;
the components
$
g_{pq}(g;R)
$
are presented in Part II by the list (2.60)-(2.61).
The $\cE_g^{PD}$-metric tensor (5.6)
can also be obtained as  the transform
\be
G_{pq}(g;
R,S)
=
\si^r_p(g;R)
\si^s_q(g;S)
n_{rs}(g;
{\bf t}_1,
{\bf t}_2)
\ee
(cf. Eq. (3.47) in Part II)
of the two-vector quasi-euclidean tensor (2.2),
where
$$
t^r_1=\si^r(g;R),\qquad
t^s_2=\si^s(g;S)
$$
(cf. Eqs. (3.10) and (3.35) in Part II).

At equal vectors the two-vector scalar product
(5.1) is  exactly  the squared Finslerian metric function:
\be
<R,R>=
K^2(g;R),
\ee
where
$
K(g;R)
$
is the function given in Part II by Eq. (2.30).

In the original
$\cE^{PD}_g$-space,
the general solution to the geodesic equation (1.1) reads
\be
R^p(s)=\mu^p(g;t(g;s)),
\ee
where $
t(g;s)$
is given by (1.35), and
$
\mu^p
$
are the functions which
realize the quasi-euclidean transformation according to Eqs.
(3.14)-(3.15) of Part II.

Particularly, from (5.2) it directly ensues that  the angle value
$\al$ of a vector $R$ with the $R^N$-axis is equal to \be \al =
\fr1h\arccos \fr { A(g;R) } { \sqrt{B(g;R)}\ } \ee and with
respect to the $\{{\bf R}\}$-plane is equal to \be \al =
\fr1h\arccos \fr { L(g;R) } { \sqrt{B(g;R)}\ }; \ee here, $B, L$,
and $A$ are respectively the functions (2.30), (2.36), and (2.39)
of Part II.

\ses
\ses
\ses
\ses

{\bf\large PART II:}

\ses
\ses

{\bf\large FINSLEROID--SPACE  $\cE^{PD}_g$  OF  POSITIVE--DEFINITE  TYPE}
\bigskip

\ses
\ses

\setcounter{sctn}{1}
\setcounter{equation}{0}

{\nin\bf 2.1. Motivation}

\ses
\ses

Below
Sec. 2.2 presents the notation and the conventions for
the space
 $\cE^{PD}_g$
and introduces necessary initial concepts and definitions.
The space
 $\cE^{PD}_g$
is constructed by assuming an axial symmetry and, therefore, the space
incorporates a single preferred direction, which we shall often refer as
the $Z$-axis.
The abbreviations FMF and FMT will be used for the Finslerian
metric function and the Finslerian metric tensor, respectively.

A characteristic parameter $g$ may take on the values between $-2$ and $2$;
at $g=0$ the space
 $\cE^{PD}_g$
is reduced to become an ordinary euclidean one. After preliminary
introducing a characteristic quadratic form $B$, which is distinct
from the euclidean sum of squares by occurrence of a mixed term
(see Eq. (2.22)), we define the FMF $K$ for the space $\cE^{PD}_g$
by the help of the formulae (2.30)-(2.33). A characteristic
feature of the formulae is the occurence of the function
$``\arctan"$. Next, we calculate basic tensor quantities of the
space. There appears a remarkable phenomenon, which essentially
simplify all the constructions, that the associated Cartan tensor
proves to be of a simple algebraic structure (see Eqs.
(2.66)-(2.67)). In particular, the phenomenon gives rise to a
conclusion that the indicatrix (the extension of the sphere) of
the space $\cE^{PD}_g$ is a space of constant curvature. The value
of the curvature depends on the parameter $g$ according to the law
(2.73).

Sec. 2.3 introduces the idea of quasi-euclidean map for the $
\cE^{PD}_g$-space. The idea is fruitful in that the
quasi-euclidean space is simple in many aspects, so that the
relevant transformation makes reduce various calculations. Last
Sec. 2.4 offers nearest interesting properties of the
quasi-euclidean metric tensor.

\ses
\ses

\setcounter{sctn}{2}
\setcounter{equation}{0}

{\nin\bf 2.2. Initial items}

\ses
\ses

Suppose we are given an
$N$--dimensional vector space $V_N$. Denote by $R$ the vectors constituting
the space, so that $R\in V_N$. Any given vector $R$ assigns a particular
direction in $V_N$. Let us fix a member
$R_{(N)}\in V_N$, introduce the
straightline
$e_N$
oriented along the vector
$R_{(N)}$,
 and use this
$e_N$
to serve as a $R^N$--coordinate axis
in $V_N$.
In this way we get the topological product
\be
V_N=          V_{N-1}   \times e_N
\ee
together with the separation
\be
R=\{\bR,R^N\}, \qquad R^N\in e_N \quad {\rm and} \quad \bR\in V_{N-1}.
\ee
For convenience, we shall frequently use the notation
\be
R^N=Z
\ee
and
\be
R=\{\bR,Z\}.
\ee
Also, we introduce a euclidean metric
\be
q=q(\bR)
\ee
over the $(N-1)$--dimensional vector space
$V_{N-1}$.

With respect to an admissible coordinate basis $\{e_a\}$ in
$V_{N-1}$, we obtain the coordinate representations \be
\bR=\{R^a\}=\{R^1,\dots,R^{N-1}\} \ee and \be
R=\{R^p\}=\{R^a,R^N\}\equiv\{R^a,Z\}, \ee together with \be
q(\bR)=\sqrt{r_{ab}R^aR^b}, \ee where $r_{ab}$ are the components
of a symmetric positive--definite tensor defined over $V_{N-1}$.
The indices $(a,b,\dots)$ and $(p,q,\dots)$ will be specified over
the ranges $(1,\dots,N-1)$ and $(1,\dots,N)$, respectively; vector
indices are up, co--vector indices are down; repeated up--down
indices are automatically summed; the notation $\de^a_b$ will
stand for the Kronecker symbol. The variables \be w^a=R^a/Z, \quad
w_a=r_{ab}w^b, \quad w= q/Z, \ee where \be w\in(-\iy,\iy), \ee are
convenient whenever $Z\ne0$. Sometimes we shall mention the
associated metric tensor \be r_{pq}=\{r_{NN}=1,~r_{Na}=0,~r_{ab}\}
\ee meaningful over the whole vector space $V_N$.

Given a parameter $g$ subject to the inequality
\be
-2<g<2,
\ee
we introduce the convenient notation
\be
h=\sqrt{1-\fr14g^2},
\ee
\ses
\be
G=g/h,
\ee
\ses
\be
g_+=\fr12g+h, \qquad g_-=\fr12g-h,
\ee
\medskip
\be
g^+=-\fr12g+h, \qquad g^-=-\fr12g-h,
\ee
so that
\be
 g_++g_-=g, \qquad g_+-g_-=2h,
\ee
\medskip
\be
 g^++g^-=-g, \qquad g^+-g^-=2h,
\ee
\ses
\be
(g_+)^2+(g_-)^2=2,
\ee
\ses
\be
(g^+)^2+(g^-)^2=2,
\ee
and
\be
g_+\g -g_-, \qquad g^+\g -g^-.
\ee

{\it The characteristic
quadratic form}
\be
B(g;R)=Z^2+gqZ+q^2
\equiv\fr12\Bigl[(Z+g_+q)^2+(Z+g_-q)^2\Bigr]>0
\ee
is of the negative discriminant, namely
\be
D_{\{B\}}=-4h^2<0,
\ee
because of Eqs. (2.12) and (2.13).
Whenever $Z\ne0$, it is also convenient to use the quadratic form
\be
Q(g;w)\eqdef B/(Z)^2,
\ee
obtaining
\be
Q(g;w)=1+gw+w^2>0,
\ee
together with the function
\be
E(g;w)
\eqdef
1+\fr12gw.
\ee
The identity
\be
E^2+h^2w^2=Q
\ee
can readily be verified.
In the limit $g\to 0$,
the definition (2.22) degenerates to the
 quadratic form of the input metric tensor (2.11):
\be
B|_{_{g=0}}=r_{pq}R^pR^q.
\ee
Also
\be
Q|_{_{g=0}}=1+w^2.
\ee

In terms of this notation, we propose the FMF
\be
K(g;R)=
\sqrt{B(g;R)}\,J(g;R),
\ee
where
\be
J(g;R)=\e^{\frac12G\Phi(g;R)},
\ee
\medskip
\medskip
\be
\Phi(g;R)=
\fr{\pi}2+\arctan \fr G2-\arctan\Bigl(\fr{q}{hZ}+\fr G2\Bigr),
\qquad {\rm if} \quad Z\ge 0,
\ee
\medskip
\be
\Phi(g;R)=
-\fr{\pi}2+\arctan \fr G2-\arctan\Bigl(\fr{q}{hZ}+\fr G2\Bigr),
\qquad  {\rm if}  \quad Z\le 0,
\ee
\ses\\
or in other convenient forms,
\ses\\
\be
\Phi(g;R)=
\fr{\pi}2+\arctan \fr G2-\arctan\Bigl(\fr{L(g;R)}{hZ}\Bigr),
\qquad  {\rm if}  \quad Z\ge 0,
\ee
\medskip
\be
\Phi(g;R)=
-\fr{\pi}2+\arctan \fr G2-\arctan\Bigl(\fr{L(g;R)}{hZ}\Bigr),
\qquad  {\rm if}  \quad Z\le 0,
\ee
where
\be
L(g;R)
=q+\fr g2Z,
\ee
and
\be
\Phi(g;R)=
\fr{\pi}2
-\arctan{\fr{hq}{A(g;R)}},
\qquad  {\rm if}  \quad Z\ge 0,
\ee
\medskip
\be
\Phi(g;R)=
-\fr{\pi}2
-\arctan{\fr{hq}{A(g;R)}},
\qquad  {\rm if}  \quad Z\le 0,
\ee
\ses\\
where
\be
A(g;R)=Z+\fr12gq.
\ee
\ses\\
This FMF has been normalized to show the handy properties
\ses\\
\be
-\fr{\pi}2
\le\Phi\le
\fr{\pi}2,
\ee
\medskip
\medskip
\be
\Phi=
\fr{\pi}2,
\quad {\rm if} \quad q=0 \quad {\rm and} \quad Z>0;
\qquad
\Phi=
-\fr{\pi}2,\quad {\rm if} \quad q=0 \quad {\rm and} \quad Z<0.
\ee

We also have
\ses\\
\be
\cot\Phi=\fr{hq}A, \qquad
\Phi|_{_{Z=0}}=
\arctan \fr G2.
\ee
\ses\\
It is often convenient to use the indicator of sign
$
\epsilon_Z
$
 for the argument $Z$:
\ses
\ses
\be
\epsilon_Z=1, \quad {\rm if}\quad Z>0;
\qquad
\epsilon_Z=-1, \quad {\rm if}\quad Z<0;
\ee

\ses

Under these conditions, we call the considered space {\it the
$\cE^{PD}_g$--space}:
\be
\cE^{PD}_g=\{V_N=V_{N-1}\times e_N;\,R\in V_N;\,K(g;R);\,g\}.
\ee

The right--hand part of the definition (2.30)
can be considered to be a function
$\breve K$
of  the arguments
$\{g;q,Z\}$,
such that
\be
\breve K(g;q,Z)
=K(g;R).
\ee
We observe that
\be
\breve K(g;q,-Z)\ne \breve K(g;q,Z),\qquad unless \quad g=0.
\ee
Instead, the function $\breve K$ shows the property of
$gZ$--\it parity\rm
\be
\breve K(-g;q,-Z)=\breve K(g;q,Z).
\ee
The $(N-1)$--space reflection invariance
holds true
\be
 K(g;R)\stackrel{R^a\leftrightarrow -R^a}{\Leftrightarrow} K(g;R).
\ee

It is frequently convenient to rewrite the representation (2.30) in the form
\be
K(g;R)=|Z|V(g;w),
\ee
whenever $Z\ne0$, with {\it the generating metric
function}
\be
V(g;w)=\sqrt{Q(g;w)}\,
j(g;w).
\ee
We have
$$
j(g;w)=J(g;1,w).
$$
Using (2.25) and (2.31)--(2.35), we obtain
\be
V'=wV/Q,
\qquad
V''=V/Q^2,
\ee
\ses
\be
(V^2/Q)'=-gV^2/Q^2,  \qquad (V^2/Q^2)'=-2(g+w)V^2/Q^3,
\ee
\ses
\be
j'=-\fr12gj/Q,
\ee
\bigskip\\
and also
\be
\fr12(V^2)'=wV^2/Q,
\qquad\quad
\fr12(V^2)''=(Q-gw)V^2/Q^2,
\ee
\ses
\be
\fr14(V^2)'''=-gV^2/Q^3,
\ee
together with
\be
\Phi'=-h/Q,
\ee
where the prime ($'$) denotes the differentiation with respect to~$w$.

\ses

Also,
\addtocounter{equation}{1}
$$
(A(g;R))^2+h^2q^2=B(g;R)
\eqno(\theequation{a})
$$
and
$$
(L(g;R))^2+h^2Z^2=B(g;R).
\eqno(\theequation{b})
$$
\ses

Sometimes it is convenient to use the function
\be
E(g;w)\eqdef 1+\fr12gw.
\ee

The simple results for these derivatives reduce
the task of computing the components of the associated FMT
to an easy exercise, indeed:
$$
R_p\eqdef\fr12\D{K^2(g;R)}{R^p}:
$$
\ses\ses
\be
R_a=
r_{ab}R^b
\fr{K^2}{B},
\qquad
R_N=(Z+gq)
\fr{K^2}{B};
\ee
\ses
\ses
\ses
\ses
$$
g_{pq}(g;R)
\eqdef\fr12\,
\fr{\prtl^2K^2(g;R)}{\prtl R^p\prtl R^q}
=\fr{\prtl R_p(g;R)}{\prtl R^q}:
$$
\ses
\ses
\ses
\ses
\be
 g_{NN}(g;R)=[(Z+gq)^2+q^2]
\fr{K^2}{B^2},
\qquad
g_{Na}(g;R)=gq
r_{ab}R^b
\fr{K^2}{B^2},
\ee
\ses
\ses
\be
g_{ab}(g;R)=
\fr{K^2}{B}
r_{ab}-g\fr{
r_{ad}R^d
r_{be}R^e
Z
}{q}
\fr{K^2}{B^2}.
\ee
\ses\\
The reciprocal tensor components are
\ses
\be
g^{NN}(g;R)=(Z^2+q^2)
\fr1{K^2},
\qquad
g^{Na}(g;R)=-gqR^a
\fr1{K^2},
\ee
\ses
\be
g^{ab}(g;R)=
\fr{B}{K^2}
r^{ab}+g(Z+gq)\fr{R^aR^b}{q}
\fr1{K^2}.
\ee
\ses\\
The determinant of the FMT
given by Eqs. (2.59)--(2.60) can readily be found in the form
\ses
\be
\det(g_{pq}(g;R))=[J(g;R)]^{2N}\det(r_{ab})
\ee
\ses\\
which shows, on noting (2.31)--(2.33), that
\ses
\be
\det(g_{pq})>0 {\it \quad over~all~the~definition~range} \quad
V_N\setminus 0.
\ee

The associated angular metric tensor
$$
h_{pq}\eqdef g_{pq}-R_pR_q\fr1{K^2}
$$
proves to be given by the components
$$
h_{NN}(g;R)=q^2
\fr{K^2}{B^2},
\qquad h_{Na}(g;R)=-Z
r_{ab}R^b
\fr{K^2}{B^2},
$$
\ses
$$
h_{ab}(g;R)=
\fr{K^2}{B}
r_{ab}-(gZ+q)\fr{
r_{ad}R^d
r_{be}R^e
}q
\fr{K^2}{B^2},
$$
\ses\\
which entails
$$
\det(h_{ab})=\det(g_{pq})\fr1{V^2}.
$$

The  use of the components of the Cartan tensor (given explicitly
in the end of the present section)
 leads,
after rather tedious straightforward calculations, to
the following simple and remarkable result.
\ses

\bf PROPOSITION 1. \it The Cartan tensor associated with the FMF
\rm(2.30) \it is of the following special
algebraic form:
\be
C_{pqr}=\fr1N\lf(h_{pq}C_r+h_{pr}C_q+h_{qr}C_p-\fr1{C_sC^s}C_pC_qC_r\rg)
\ee
with
\be
C_tC^t=\fr{N^2}{4K^2}g^2.
\ee
\rm
\ses

By the help of (2.65), elucidating the structure of
the curvature tensor
\be
S_{pqrs}\eqdef(C_{tqr}\3Cpts-C_{tqs}\3Cptr)
\ee
results in the simple representation
\be
S_{pqrs}=-\fr{C_tC^t}{N^2}(h_{pr}h_{qs}-h_{ps}h_{qr}).
\ee
Inserting here (2.66), we are led to
\ses

\bf PROPOSITION 2. \it The curvature tensor of the space
$\cE^{PD}_g$ is of the special type
\be
S_{pqrs}=S^*(h_{pr}h_{qs}-h_{ps}h_{qr})/K^2
\ee
with \rm
\be
S^*=-\fr14g^2.
\ee
\ses

{\bf DEFINITION}.\, FMF (2.30) introduces an $(N-1)$--dimensional
indicatrix hypersurface  according to the equation
\be
K(g;R)=1.
\ee
We call this particular hypersurface \it the Finsleroid\rm,
to be denoted as
$
\cF^{PD}_g.
$

Recalling the known formula
$
\cR=1+S^*
$
for the indicatrix curvature
(see [5]),
from (2.71) we conclude that
\be
\cR_{Finsleroid}=h^2=1-\fr14g^2, \qquad\quad 0 < \cR_{Finsleroid} \le 1.
\ee
 Geometrically, the fact that the quantity ~(2.70)
is independent of  vectors~$R$ means that the
indicatrix curvature is  constant. Therefore, we have arrived at
\ses

\bf PROPOSITION 3. \it The Finsleroid
$
\cF^{PD}_g
$
is a constant-curvature space with the
positive curvature value~\rm(2.73).
\ses

Also, on comparing between the result (2.73) and  Eqs. (2.22)--(2.23), we obtain
\ses

\bf PROPOSITION 4. \it The Finsleroid curvature  relates to
the discriminant
\rm (2.23)
\it
of the
input characteristic quadratic form~\rm (2.22)
\it
simply as
\be
\cR_{Finsleroid}=-\fr14D_{\{B\}}.
\ee
\rm

Last, we write down the explicit components of the relevant Cartan tensor
\ses\\
$$
C_{pqr}\eqdef \fr12\D{g_{pq}}{R^r}:
$$
\ses
\ses
$$
R^NC_{NNN}=gw^3V^2Q^{-3}, \quad\quad R^NC_{aNN}=-gww_aV^2Q^{-3},
$$
\ses
$$
R^NC_{abN}=\fr12gwV^2Q^{-2}r_{ab}+\fr12g(1-gw-w^2)w_aw_bw^{-1}V^2Q^{-3},
$$
\ses
$$
R^NC_{abc}= -\fr12gV^2Q^{-2}w^{-1}(r_{ab}w_c+r_{ac}w_b+r_{bc}w_a)
 +gw_aw_bw_cw^{-3}\lf(\fr12Q+gw+w^2\rg)V^2Q^{-3};
$$
\ses\\
and
\ses\\
$$
R^N\3CNNN=gw^3/Q^2, \quad\quad\quad R^N\3CaNN=-gww_a/Q^2,
$$
\ses
$$
R^N\3CNaN=-gw(1+gw)w^a/Q^2,
$$
\ses
$$
R^N\3CaNb=\fr12gwr_{ab}/Q+\fr12g(1-gw-w^2)w_aw_b/wQ^2,
$$
\ses
$$
R^N\3CNab=\fr12gw\de_b^a/Q+\fr12g(1+gw-w^2)w^aw_b/wQ^2,
$$
\ses
$$
R^N\3Cabc=  -\fr12g
\lf(\de_a^bw_c+\de_c^bw_a+(1+gw)r_{ac}w^b
\rg)
/wQ
 +\fr12g(gwQ+Q+2w^2)w_aw^bw_c/w^3Q^2.
$$
\ses\\
The components have been calculated by the help of the formulae (2.50)--(2.53).
\ses

The use of the contractions
$$
R^N\3Cabcr^{ac}=-g\fr{w^b}w\fr{1+gw}Q\lf(\fr{N-2}2+\fr1Q\rg)
$$
and
\ses\\
$$
R^N\3Cabcw^aw^c=-g\fr{w}{Q^2}(1+gw)w^b
$$
is handy in many calculations.

Also,
\ses\\
$$
R^NC_N=\fr N2gwQ^{-1}, \quad\quad R^NC_a=-\fr N2g(w_a/w)Q^{-1},
$$
\ses
$$
R^NC^N=\fr N2gw/V^2, \quad\quad R^NC^a=-\fr N2gw^a(1+gw)/wV^2,
$$
\ses
$$
C^N=\fr N2gwR^NK^{-2}, \qquad C^a=-\fr N2gw^a(1+gw)w^{-1}R^NK^{-2},
$$
\ses
\ses
\ses
$$
C_pC^p=\fr{N^2}{4K^2}g^2.
$$

\ses
\ses

\setcounter{sctn}{3}
\setcounter{equation}{0}

{\bf 2.3. Quasi-euclidean map of Finsleroid }

\ses
\ses

It is possible to indicate
 the diffeomorphism
\be
\cF_g^{PD}
\stackrel{{\it i_g}}{\Longrightarrow}\cS^{PD}
\ee
of the Finsleroid
$\cF_g^{PD}\subset V_N$
to the unit sphere
$\cS^{PD}\subset V_N$:
\be
\cS^{PD}=\{R\in\cS^{PD}:\; S(R)=1\},
\ee
where
\be
S(R)=\sqrt{r_{pq}R^pR^q}
\equiv
\sqrt{(R^N)^2+r_{ab}R^aR^b}
\ee
is the input euclidean metric function (see (2.11)).

The
diffeomorphism (3.1)
can always be extended to get the diffeomorphic map
\be
V_N
\stackrel{\si_g}{\Longrightarrow}V_N
\ee
of the whole vector space
$V_N$
by means of the
homogeneity:
\be
\si_g\cdot (bR)=b\si_g\cdot R, \quad b>0.
\ee
To this end it is sufficient to take merely
\be
\si_g\cdot R=||R|| i_g\cdot\Bigl(\fr{R}{||R||}\Bigr),
\ee
where
\be
||R||=K(g;R).
\ee
Eqs. (3.1)--(3.7) entail
\be
K(g;R)=S(\si_g\cdot R).
\ee

The
identity (2.57) suggests to take the map
\be
\bar R=\si_g\cdot R
\ee
by means of the components
\be
 \bar R^p=\si^p(g;R)
\ee
with
\ses
\be
\si^a=
R^ah
J(g;R),
\qquad
\si^N=A(g;R)J(g;R),
\ee
\ses\\
where $J(g;R)$ and $A(g;R)$ are the functions (2.31) and (2.39).
Indeed, inserting (3.11) in (3.3) and taking into account
Eqs. (2.30) and (2.57),
we get
the  identity
\be
S(\bar R)=K(g;R)
\ee
which is tantamount to the implied relation
(3.8).

\ses

{\bf PROPOSITION 5.}  The map given explicitly by Eqs. (3.9)--(3.11)
 assigns \it the diffeomorphism between the Finsleroid and the unit sphere
\rm according to Eqs. (3.1)--(3.8).

\ses

Therefore, we may also call the operation (3.1)
{
\it
the quasi-euclidean map of Finsleroid.}

The inverse
\ses
\be
 R=\mu_g\cdot \bar R,
  \qquad \mu_g=(\si_g)^{-1},
\ee
\ses\\
of the transformation (3.9)--(3.11)
can be presented by the components
\be
  R^p=\mu^p(g;\bar R)
\ee
with
\be
\mu^a=\bar R^a/hk(g;\bar R),
\qquad
\mu^N=I(g;\bar R)/k(g;\bar R),
\ee
\ses\\
where
\ses
\be
k(g;\bar R)\eqdef J(g;\mu(g;\bar R))
\ee
\ses\\
and
\be
I(g;\bar R)
=\bar R^N-\frac12G\sqrt{r_{ab}\bar R^a\bar R^b}.
\ee
The identity
\ses
\be
\mu^p(g;\si(g;R))\equiv R^p
\ee
\ses\\
can readily be verified.
Notice that
\ses
\be
\fr{\sqrt{r_{ab}\bar R^a\bar R^b}}{\bar R^N}
=\fr{hq}{A(g;R)},
\qquad
w^a=\fr{R^a}{R^N}=\fr{\bar R^a}{hI(g;\bar R)},
\ee
\ses\\
and
\be
\sqrt{B}/Z=S/I,
\qquad
\sqrt{Q}=S/I.
\ee

The $\si_g$--image
\ses
\be
\phi(g;\bar R)
\eqdef\Phi(g;R)|_{_{R=\mu(g;\bar R)}}
\ee
\ses\\
of the function $\Phi$ described by Eqs. (2.31)--(2.42)
is of a clear meaning of  angle:
\ses
\ses
\be
\phi(g;\bar R)=\arccot\fr{\bar R^N}{\sqrt{r_{ab}\bar R^a\bar R^b}}
=\left\{
\ba{rcl}
\fr{\pi}2-\arctan\fr{\sqrt{r_{ab}\bar R^a\bar R^b}}{\bar R^N},
\quad {\rm if}\quad \bar R^N\ge0;\\
\ses\\
-\fr{\pi}2-\arctan\fr{\sqrt{r_{ab}\bar R^a\bar R^b}}{\bar R^N},
\quad {\rm if}\quad\bar R^N\le0;\\
\ea
\right.
\ee
\ses\\
which ranges over
\ses
\ses
\be
-\fr{\pi}2
\le\phi\le
\fr{\pi}2.
\ee
\ses\\
We have
\ses
\be
\phi=
\fr{\pi}2,
\quad {\rm if} \quad \bar R^a=0 \quad {\rm and} \quad \bar R^N>0;
\qquad
\phi=
-\fr{\pi}2
,\quad {\rm if} \quad \bar R^a=0 \quad {\rm and} \quad \bar R^N<0,
\ee
\ses\\
and also
\ses
\be
\phi|_{_{\bar R^N=0}}=0.
\ee
\ses\\
Comparing Eqs. (3.16) and (2.31) shows that
\ses
\be
k=\e^{\frac12G\phi}.
\ee

The right--hand parts in
(3.11) are homogeneous
 functions
of degree~1:
\be
\si^p(g;bR)=b\si^p(g;R), \quad b>0.
\ee
Therefore, the identity
\be
\si_s^p(g;R)R^s=\bar R^p
\ee
should be valid for the derivatives
\ses\\
\be
\si_p^q(g;R)\eqdef\D{\si^q(g;R)}{R^p}.
\ee
The simple representations
\be
\si^N_N(g;R) =\lf(B+\fr12gqA\rg)\fr{J}{B},
\ee
\bigskip
\be
\si^N_a(g;R) =-\fr{g(ZA-B)}{2q}\fr{J r_{ab}R^b}{B},
\ee
\bigskip
\be
\si^a_N(g;R) =\fr12gq\fr{J R^ah}{B},
\ee
\bigskip
\be
\si^a_b(g;R) =\lf(B\delta^a_b-
\fr{gr_{bc}R^cR^aZ}{2q}\rg)\fr{Jh}{B},
\ee
\ses\\
and also the determinant
value
\ses
\be
\det(\si_p^q)=h^{N-1}J^N
\ee
\ses\\
are obtained. The relations
$$
\si_b^aR^b=JhR^a(AZ+q^2)/B, \qquad
r^{cd}\si_c^a\si_d^b=
J^2h^2\lf[r^{ab}-g(R^aR^bZ/qB)+\fr14g^2(R^aR^bZ^2/B^2)\rg]
$$
are handy in many calculations involving the coefficients $\{\si_p^q\}$.

Henceforth, to simplify notation, we shall use the substitution
\be
t^p=\bar R^p.
\ee

Again, we can note the homogeneity
\be
\mu^p(g;bt)=b\mu^p(g;t), \quad b>0,
\ee
for the functions (3.15), which entails the identity
\be
\mu_s^p(g;t)t^s=R^p
\ee
for the derivatives
\ses
\be
\mu_q^p(g;t)\eqdef\D{\mu^p(g;t)}{t^q}.
\ee
\ses\\
We find
\ses
\be
\mu_N^N=1/k(g;t)-\fr12g
\fr{m(t)
I(g;t)}{k(g;t)(S(t))^2},
\qquad
\mu_a^N=\fr12g\fr{r_{ac}t^cI^*(g;t)}{k(g;t)(S(t))^2},
\ee
\ses
\ses
\be
\mu_N^a=-\fr12g
\fr{m(t)
\,t^a}{hk(g;t)(S(t))^2},
\qquad
\mu_b^a=\fr1{hk(g;t)}\de_b^a+
\fr12g\fr{t^Nt^ar_{bc}t^c}
{m(t)
\,hk(g;t)(S(t))^2},
\ee
\ses\\
where
\be
m(t)=
\sqrt{r_{ab}t^at^b},
\ee
\ses
\be
I^*(g;t)=hm(t)-\fr12gt^N,
\ee
and
\ses
\be
S(t)=\sqrt{r_{rs}t^rt^s}
\equiv
\sqrt{(t^N)^2+r_{ab}t^at^b}.
\ee

\ses
The  relations
\ses
$$
\D{(1/k(g;t))}{t^N}=-\fr12g\fr{m(t)}{hk(g;t)(S(t))^2},
\quad\quad
\D{(1/k(g;t))}{t^a}=\fr12g\fr{t^Nr_{ab}t^b}
{m(t)hk(g;t)(S(t))^2}
$$
\ses\\
are obtained.

Also
\be
 R_p\mu^p_q=t_q,
\qquad
t_p\si^p_q=R_q.
\ee

The unit vectors
\be
L^p\eqdef\fr{t^p}{S(t)}, \qquad L_p\eqdef r_{pq}L^q
\ee
fulfil the relations
\be
L^q=l^p\si^q_p,
\qquad l^p=\mu^p_qL^q,
\qquad l_p=\si^q_pL_q,
\qquad L_p=\mu^q_pl_q,
\ee
where $l^p=R^p/K(g;R)$ and $l_p=g_{pq}(g;R)l^q$ are
the initial Finslerian unit vectors.

Now we use the explicit formulae (2.61)--(2.62) and (3.29)--(3.32) to
find the transform
\be
n^{rs}(g;t)\eqdef \si_p^r\si_q^sg^{pq}
\ee
of
the FMT $g_{pq}$
under the $\cF_g^{PD}$--induced map (3.9)--(3.11), which results in
\ses

{\bf PROPOSITION 6.} \it One obtains the simple representation
\ses

\be
n^{rs}=h^2r^{rs}+\fr14g^2L^rL^s.
\ee
The covariant version reads
\be
n_{rs}=
\fr1{h^2}r_{rs}-\fr14G^2L_rL_s.
\ee
The determinant of this tensor is a constant:\rm
\be
\det(n_{rs})=h^{2(1-N)}\det(r_{ab}).
\ee
\bigskip

Notice that
$$
L^pL_p=1, \quad n_{pq}L^q=L_p, \quad n^{pq}L_q=L^p, \quad
n_{pq}L^pL^q=1, \quad n_{pq}t^pt^q=(S(t))^2.
$$
\ses

Eq. (5.47) obviously entails
\be
g_{pq}=
n_{rs}(g;t)\si_p^r\si_q^s.
\ee

\ses
\ses

\setcounter{sctn}{4}
\setcounter{equation}{0}

{\bf 2.4.  Quasi-euclidean metric tensor}

\ses
\ses

Let us introduce
\ses

{\bf DEFINITION}.  The metric tensor (3.48)--(3.49) is called {\it
quasi-euclidean}. \ses

\ses

{\bf DEFINITION}.  \it
 The quasi-euclidean space
\be
{\cal Q}_N=\{V_N;n_{pq}(g;t);g\}
\ee
\rm
is an extension of the euclidean space $\{V_N;r_{pq}\}$ to the case $g\ne0$.
\ses

The transformation (3.47) can be inverted to read \be
g_{pq}=\si_p^r\si_q^sn_{rs}. \ee For the angular metric tensor
(see the formula going below Eq. (2.64)), from (3.46) and (4.2) we
infer \be h_{pq}=\si_p^r\si_q^sH_{rs}\fr1{h^2}, \ee where \be
H_{rs}\eqdef r_{rs}-L_rL_s \ee is the tensor showing the
orthogonality property \be L^rH_{rs}=0. \ee \ses One can readily
find that
$$
H_{rs}= h^2(n_{rs}-L_rL_s).
$$

\ses

{\bf PROPOSITION  7}.  The quasi-euclidean metric tensor
\rm(3.48)--(3.49) {\it is conformal} to the  euclidean metric
tensor. \ses

Indeed, if we consider the map
\be
\bar R^p\rightarrow \tilde R:\qquad
\tilde R^p=f(g;\bar R)\bar R^p/h
\ee
with
\be
f(g;\bar R)=a\lf(g;\,\fr12S^2(\bar R)\rg)
\ee
and use the coefficients
\be
k_q^p\eqdef \D{\tilde R^p}{\bar R^q}
=(f\de_q^p+a'\bar R^p\bar R_q)/h
\ee
to define the tensor
\bigskip
\be
c^{pq}(g;\tilde R)\eqdef k_r^pk_s^qn^{rs}(g;\bar R),
\ee
we find that
\be
c^{pq}=f^2r^{pq}
\ee
whenever
\be
f=\lf[\fr12S^2(\bar R)\rg]^{\ga/2},
\ee
where
\be
\ga=h-1\equiv \sqrt{1-\fr{g^2}4}-1
\ee
 is the parameter. The proof of Proposition 7
is complete.

Let us now use the obtained quasi-euclidean metric tensor $
n_{pq}(g;t)$ to construct the associated {\it quasi-euclidean
Christoffel symbols} $\3N prq(g;t)$. We find consecutively: \be
n_{pq,r}\eqdef\D{n_{pq}}{t^r}=-\fr14G^2(H_{pr}L_q+H_{qr}L_p)/S,
\ee and \be \3N prq=n^{rs}N_{psq}, \qquad
N_{prq}=\fr12(n_{pr,q}+n_{qr,p}-n_{pq,r}), \ee together with \be
N_{prq}(g;t)=-\fr14G^2H_{pq}L_r/S, \ee which eventually yields \be
\3N prq(g;t)=-\fr14G^2L^rH_{pq}/S. \ee Comparing the
representation (4.16) with the identity (4.5) shows that \be
t^p\3N prq=0, \qquad \3N pss=0,\qquad \3 N tsr \3N ptq=0. \ee

Also,
\be
\D{\3N prq}{t^s}-\D{\3N prs}{t^q}
 =-\fr14G^2(H_{pq}{H_s}^r-H_{ps}{H_q}^r)/S^2.
\ee
Using the identities (4.17)-(4.18) in
{\it the quasi-euclidean  curvature tensor}:
\be
R_p{}^r{}_{qs}(g;t)\eqdef\D{N_p{}^r{}_q}{t^s}-\D{N_p{}^r{}_s}{t^q}+
N_p{}^w{}_qN_w{}^r{}_s-N_p{}^w{}_sN_w{}^r{}_q,
\ee
we arrive at the simple result:
\bigskip
\be
R_{prqs}(g;t)     =-\fr14G^2(H_{pq}H_{rs}-H_{ps}H_{qr})/S^2.
\ee
This infers the identities
\be
L^pR_{pqrs}=L^qR_{pqrs}=L^rR_{pqrs}=L^sR_{pqrs}=0.
\ee
\ses

{\bf NOTE}. Because of the transformation rules (3.12) and (3.47),
the representation (4.20) is tantamount to
Eqs. (2.69)--(2.70).
Therefore {\it we have got another rigorous proof
of Proposition } 3, {\it and of
Eq.} (2.71), {\it concerning the Finsleroid curvature}.

\bigskip
\bigskip

\def\bibit[#1]#2\par{\rm\noindent\parskip1pt
                     \parbox[t]{.05\textwidth}{\mbox{}\hfill[#1]}\hfill
                     \parbox[t]{.925\textwidth}{\baselineskip11pt#2}\par}

\nin{\bf References}
\bigskip

\bibit[1] E. Cartan: \it Les espaces de Finsler, Actualites \rm 79,
Hermann,
Paris 1934.

\bibit[2] H. Busemann: \it Canad. J. Math. \bf1 \rm(1949), 279.

\bibit[3] H. Rund: \it The Differential Geometry of Finsler spaces, \rm
Springer-Verlag, Berlin 1959.

\bibit[4] R. S. Ingarden: \it Tensor \bf30 \rm(1976), 201.

\bibit[5] G. S. Asanov: \it Finsler Geometry, Relativity and Gauge Theories, \rm
D.~Reidel Publ. Comp., Dordrecht 1985.

\bibit[6] D.~Bao, S. S. Chern, and Z. Shen (eds.): \it Finsler Geometry \rm
(Contemporary Mathematics, v.~196), American Math. Soc., Providence 1996.

\bibit[7] D.~Bao, S. S. Chern, and Z. Shen: \it An
Introduction to Riemann-Finsler Geometry,
\quad\rm Springer, N.\.Y., Berlin, 2000.

\bibit[8] A.C. Thompson: \it Minkowski Geometry,\quad\rm Cambridge
University Press, Cambridge. 1996.

\bibit[9] G. S. Asanov: \it Aeq. Math. \bf49 \rm(1995), 234.

\bibit[10] G.S. Asanov:
arXiv:hep-ph/0306023, 2003.

\bibit[11] G. S. Asanov: \it Rep. Math. Phys. \bf 45 \rm(2000), 155;
\bf 47 \rm(2001), 323.

\bibit[12] G.S. Asanov: \it Moscow University Physics Bulletin
 \bf49\rm(1) (1994),~18; \bf51\rm(1) (1996),~15; \bf51\rm(2)
 (1996),~6; \bf51\rm(3) (1996),~1; \bf53\rm(1) (1998),~15.

\bibit[13] C. M{\o}ller: \it The Theory of Relativity, \rm
Claredon Press, Oxford  1972.

\bibit[14] J. L. Synge: \it Relativity: The General Theory, \rm
North-Holland, Amsterdam 1960.

\end {document}